\documentclass[11pt]{article}  
\usepackage[letterpaper, left=1in, top=1in, right=1in, bottom=1in,nohead,includefoot, verbose, ignoremp]{geometry}
	\usepackage[usenames,dvipsnames,x11names]{xcolor}
\usepackage{charter,xcolor,times,graphicx,natbib,amsmath,amssymb} 
	\usepackage[pdftex,pagebackref=true]{hyperref}
	\hypersetup{
	colorlinks,
	linkcolor=RoyalBlue1,
	urlcolor=RoyalBlue2,
	citecolor=RoyalBlue3
	}

\def\para#1{\noindent{\bf#1}}
\def\eqn#1{equation~(\ref{eq:#1})}
\def\eqns#1#2{equations~(\ref{eq:#1},\ref{eq:#2})}

\def\A{\mbox{\boldmath$A$}}
\def\B{\mbox{\boldmath$B$}}

\def\U{\mbox{\boldmath$U$}}
\def\D{\mbox{\boldmath$D$}}
\def\G{\mbox{\boldmath$G$}}
\def\V{\mbox{\boldmath$V$}}

\def\zero{\mbox{\boldmath$0$}}

\def\y{\mbox{\boldmath$y$}}
\def\a{\mbox{\boldmath$a$}}

\def\v{\mbox{\boldmath$v$}}

\def\z{\mbox{\boldmath$z$}}
\def\bbeta{\mbox{\boldmath$\beta$}}

\def\bSigma{\mbox{\boldmath$\Sigma$}}

\def\btheta{\mbox{\boldmath$\theta$}}

\def\e{\mbox{\boldmath$e$}}
\def\f{\mbox{\boldmath$f$}}

\def\bnu{\mbox{\boldmath$\nu$}}

\def\bPsi{\mbox{\boldmath$\Psi$}}

\def\bdelta{\mbox{\boldmath$\delta$}}
\def\bxi{\mbox{\boldmath$\xi$}}
\def\bGamma{\mbox{\boldmath$\Gamma$}}

\begin{document}

\title{Dynamics \& Sparsity in Latent Threshold Factor Models: \break
A Study in Multivariate EEG Signal Processing}     
\author{Jouchi Nakajima\footnote{The Deputy Director of Global Economic Research,
			Bank of Japan, 2-1-1 Nihonbashi-Hongokucho, Chuo-ku, Tokyo 103-0021, Japan.
			  \href{mailto:jouchi.nakajima@gmail.com}{jouchi.nakajima@gmail.com}} 
		\,\, \& 
	    Mike West\footnote{The Arts \& Sciences Professor of Statistics \& Decision Sciences,
 					Department of Statistical Science, Duke University, Durham 27708-0251, USA.
					\href{mailto:mw@stat.duke.edu}{mw@stat.duke.edu}.
			\newline\newline
			This manuscript is the author original; revised version to appear in {\em \color{RoyalBlue2} 				The Brazilian Journal of Probability and Statistics}. 
			The research reported here was supported in part by a grant from the U.S. National Science Foundation 	(DMS-1106516).
			Any opinions, findings and conclusions or recommendations expressed in this work are 
			those of the authors and do not necessarily reflect the views of the NSF or the Bank of Japan.
			}
	  }
\def\today{}
\maketitle

\setcounter{page}{0}
\thispagestyle{empty}

\begin{abstract}
We discuss Bayesian analysis of   multivariate time series with dynamic factor models that exploit time-adaptive sparsity in model parametrizations 
via the latent threshold approach. One central focus is on the transfer responses of multiple interrelated series to underlying, dynamic latent 
factor processes. Structured priors  on model hyper-parameters are key to the efficacy of dynamic latent thresholding, and MCMC-based computation
enables model fitting and analysis.  A detailed case study of electroencephalographic (EEG) data from experimental  psychiatry highlights the use of  
latent threshold extensions of time-varying vector autoregressive and factor models.  This study explores a class  of dynamic transfer response factor 
models, extending prior Bayesian modeling of multiple EEG series and highlighting the practical utility of the latent thresholding concept in 
multivariate, non-stationary time series analysis. 

\bigskip

\bigskip

\noindent {\em Key Words \& Phrases: } Dynamic factor models;
Dynamic sparsity;
EEG time series;
Factor-augmented vector autoregression;
Impulse response;
Multivariate time series;
Sparse time-varying loadings;
Time-series decomposition;
Transfer response factor models.\\[2mm]

\end{abstract}

\clearpage
\setcounter{page}{1}

\section{Introduction}

In high-dimensional time series analysis, the need to define time-varying patterns of
sparsity in model parameters has proven challenging.   Dynamic latent thresholding, introduced
in~\cite{NakajimaWest10},  provides a general approach that induces parsimony into time series
 model structures with potential to reduce effective parameter dimension and improve model 
 interpretations as well as forecasting performance.  The utility of various classes of latent threshold models (LTMs) 
has been demonstrated in recent applied studies in  macroeconomics~\citep{NakajimaWest10,KimuraNakajima2016}
and financial forecasting and portfolio decisions~\citep{NakajimaWest11,Zhou2012}. The
scope of the approach includes dynamic regressions, dynamic latent factor models, time-varying
vector autoregressions,  and dynamic graphical models of multivariate stochastic volatility, and
also opens a path to new approaches to dynamic network modeling~\citep{Nakajima2014}.

This paper adapts the latent thresholding approach to different classes of multivariate  factor models 
with a one main interest in dynamic transfer response analysis. Our detailed case-study
concerns   time-varying lag/lead relationships
among multiple time series in electroencephalographic (EEG) studies. Here the  latent threshold analysis of such models  induces relevant, 
time-varying patterns of sparsity in otherwise time-varying factor loadings matrices, among other model features. 
We evaluate and compare two different 
classes of models in the EEG study, and explore a number of posterior  summaries in relation to this main interest.

Time series factor modeling has been an area of growth for Bayesian analysis in recent years. 
Two key themes   are: (i) dynamic factor models, where latent factors are time series processes 
underlying patterns of relationships
among multiple time 
series~\cite[e.g.][]{
Aguilar1999a,
PittShephard99,
AguilarWest00,
KoopPotter2004,
BernankeBoivinEliasz05,
LopesCarvalho07,
DelNegro2008,
KoopKorobilis10};
 and (ii) sparse factor models, where the bipartite graphs representing 
conditional dependencies of observed variables on factors are not completely 
connected~\cite[e.g.][]{
West2003,
Lucas2006,
Carvalhoetal08,
Lucas2009a,
Yoshida2010,
Carvalho11,
Bhattacharya2011}, 
increasingly applied in  problems of classification and prediction.
 
Here we combine dynamics with sparsity.  Some of the practical relevance
of models with time-varying factor loadings is evident in recent 
studies~\cite[e.g.][]{
LopesCarvalho07,
DelNegro2008,
Carvalho11}. 
As the number of variables and factors increase, so does the need to 
induce sparsity in loadings matrices to reflect the view that variables will typically be 
conditionally dependent on only a subset of factors. In a time series setting, however, the patterns of occurrence of zeros in otherwise time-varying factor loadings matrices may also
be time-varying. One factor may relate to one particular variable with a  time-varying loading over a period
of time, but be insignificant for that variable in other time periods. Thus the need to
develop models of  time-varying sparsity of loadings matrices in dynamic factor models.

\para{Conventions and notation:} All vectors are column vectors.  We use
$\y\sim N(\a,\A)$,  $d\sim U(a,b)$, $p\sim B(a,b)$, $v\sim G(a,b)$, $\U \sim W(c, \D)$,
for the normal, uniform, beta, gamma, and Wishart
distributions, respectively.  Succinct notation for ranges uses $s:t$  to denote $s,s+1,\ldots,t$ when $s<t;$
e.g., $\y_{1:T}$ denotes $\{ \y_1,\ldots,\y_T\}$. The indicator function is $I(\cdot) $  
and $\mathrm{diag}(\cdot)$ is the  diagonal matrix with diagonal elements in the argument
and hence dimension implicit.  Elements of any $c-$vector time series $\v_t$ are $v_{jt}$, $(j=1:c),$ and 
those of any $c\times d$ matrix time series $\V_t$ are $v_{ijt},$ $(i=1:c,\ j=1:d).$


\section{Dynamic Factor Models \label{sec:DFMs}} 

In a general setting, the $m-$vector  time series $\y_t = (y_{1t},\ldots, y_{mt})'$, ($t=1,2,\ldots$)  is modeled as  
	\begin{eqnarray}
	\y_t  = \A_t\z_t + \B_t \f_t + \bnu_t,     \qquad  \bnu_t  \sim N(\zero,\bSigma_t), \label{eq:DFM}
	\end{eqnarray}
where: 
\begin{itemize}\itemsep-1pt
  \item $\z_t$ is a $q-$vector of predictor variables known at time $t$; 
  \item $\A_t$ is the $m\times q$  matrix of regression coefficients at time $t$; 
  \item $\f_t$ is the $r\times 1$ vector of latent factors, arising from some underlying latent factor process over time; 
  \item $\B_t$ is the $m\times r$ matrix of  factor loadings at time $t$; 
  \item $\bnu_t$ is the residual term, assumed zero-mean normal with diagonal variance matrix
  $\bSigma_t=\mathrm{diag}(\sigma_{1:m,t}^2)$ of volatilities $\sigma_{jt}$ at time $t.$  
\end{itemize} 	
Complete specification requires models for $\f_t$,  $\A_t$, $\B_t$ and $\sigma_{jt}$ over time. 
Typically, $m>>r$, and models are identified via constraints on $\B_t$, such as fixing $\B_t$ to have zeros above a unit upper diagonal:   $b_{iit}=1$ and $b_{ikt}=0$ for $i=1:r,\ k=i+1:r.$   In  Section~\ref{sec:ModelsMandM+},  there is interpretable structure to
$\f_t$ and alternative assumptions are natural.  
Special cases and  assumptions now  follow. 

\medskip\noindent{\bf Constant and Sparse Factor Models:}  
Much past work uses constant coefficients $\A_t=\A$ and loadings $\B_t=\B.$ 
The pure factor model, with $\A_t\z_t=\zero$ and $\B_t=\B,$   typically
assumes the factors $\f_t$ are zero-mean and independent, yielding a linear factor representation of 
the conditional variance matrix of $\y_t.$  Sparsity in $\B$ then begins 
development of more parsimonious models for larger $m,r$~\citep[e.g.][]{West2003}.

\medskip\noindent{\bf FAVAR Models:} 
When $\z_t$  concatenates past values $\y_{t-j},$ $(j=1:d)$ to  lag $d,$ 
and $\A_t=\A, \B_t=\B$ are constant, the model 
is  a factor-augmented vector autoregression (FAVAR). Variants 
based on differing   models for $\f_t$ are becoming of increasing interest in
 macroeconomics~\citep{BernankeBoivinEliasz05,KoopKorobilis10}.

\medskip\noindent{\bf Factor Stochastic Volatility Models:}
Traditional Bayesian multivariate volatility models have $\A_t=\zero,$   $\B_t=\B,$
and $\f_t\sim N(\zero,\bGamma_t)$ where $\bGamma_t=\mathrm{diag}(\gamma_{1:r,t}^2).$
Model completion involves  stochastic volatility model for the   $\gamma_{jt}$ 
and $\sigma_{jt},$  based on either log-AR(1) models or Bayesian
discounting~\cite[e.g.][]{
Aguilar1999a,
PittShephard99,
AguilarWest00,
PradoWest10}.

\medskip\noindent{\bf Time-Varying Regression and Factor Loadings Models:} 
Variants of models with time-varying $\A_t, \B_t$ are well-used~\citep[e.g][]{WestHarrison97,PradoWest10,West2013}. Typically, the elements $a_{ijt}, b_{ijt}$  are AR(1) processes.  Within this class, random walk models have
flexibility to adapt to change over time, while stationary AR(1) models 
can have  longer-term predictive value and interpretation~\citep{
LopesCarvalho07,
DelNegro2008,
NakajimaWest10,
NakajimaWest11}.

\medskip\noindent{\bf Process Models for Factors:} 
Models of  factor processes $\f_t$ typically
involve either conditionally independent factors over time, with or without time-varying conditional 
variances, or stationary vector autoregressive (VAR) models.

\section{Dynamic Factor Models and Transfer Responses\label{sec:ModelsMandM+}} 

\subsection{Introductory Comments} 

We highlight  example models that incorporate elements noted in
Section~\ref{sec:DFMs}, while  being customized to the EEG study: a  
response variable is hierarchically linked to current and lagged values of an underlying 
latent process of scientific interest. A first latent factor model is discussed, then extended with a time-varying 
vector autoregressive component;  these two models are customized examples of 
time-varying FAVAR processes. 
 
\subsection{Model M:  A Dynamic Transfer Response Factor Model \label{subsec:DTRFM}}

A  dynamic transfer response factor model (DTRFM) relates the outcome variables to a foundational, 
scalar latent process $x_t$ 
by specifying $\f_t$ to be a vector of recent values of this underlying scalar process.   Each outcome variable 
relates to potentially several recent and lagged values of $x_t$ through time-varying loadings coefficients;
at any instant in time, these coefficients define the transfer response of the variable to the history of 
the underlying process. As the loadings vary in time, the form of this response then naturally varies. 
The basic structure of the model is described here. 

In \eqn{DFM}, set $\A_t\z_t=\zero$ for all $t.$ Suppose also that 
$$\f_t = (x_t,x_{t-1},\ldots,x_{t-r+1})',$$ for some $r>0,$ where the scalar series $x_t$ is modeled as 
a time-varying autoregressive (TVAR) process of order $p$. That is, 
	\begin{alignat}{4}
	x_t & = \,\, \sum_{j=1:p} \delta_{jt} x_{t-j} + \varepsilon_t, \qquad & \varepsilon_t & \sim N(0, w_t), \label{eq:xt} \\
	\bdelta_t & = \,\, \bdelta_{t-1} + \bxi_t, \qquad & \bxi_t & \sim N(\zero, \bPsi), \label{eq:delta}
	\end{alignat}
where 
$\bdelta_t = (\delta_{1t},\ldots,\delta_{pt})'$ is the vector of AR coefficients at time $t.$ Conditional on the variance elements
$w_{1:T}$ and $\bPsi$,  
 the $\bnu_t$, $\varepsilon_t$ and $\bxi_t$ sequences are assumedly independent over time and mutually independent. 
Equation~(\ref{eq:delta}) indicates that the $\bdelta_t$  coefficients  follow a vector random walk over time, 
permitting time variation but not anticipating its
direction or form.   Coupled with \eqn{xt} we have the traditional specification of 
a Bayesian TVAR$(p)$ model for the latent $x_t$ process. 

For the $i^{\rm th}$ scalar response variable, the above model implies 
	\begin{eqnarray}
	y_{it}  = \sum_{k=0:r-1} b_{ikt} x_{t-k}  + \nu_{it},     \qquad  \nu_{it}  \sim N(0,\sigma_{it}^2), \label{eq:DTRFMj}
	\end{eqnarray}
showing the transfer of responses from past values of $x_t$ via the-- possibly quite widely time-varying-- loadings $b_{ikt}$, the
latter specific to series $i,$ for each $i=1:m.$  

Model identification is straightforward. From~\eqn{DTRFMj}, it is clear that 
an identification problem exists with respect to the lag/lead structure, i.e., the time origin for the latent $x_t$ process, as
well as the scale of $x_t$ relative to the $b_{ikt}.$    Fixing elements of one row of $\B_t$ to specified values obviates this. 
Here we do this on the first row: for one factor (lag)  index $s\in \{ 1:r \},$ we set $b_{1st}=1$ and  
$b_{1kt}=0$ for $k=1:r,\ k\ne s.$   This way, $y_{1t}$ is a direct, unbiased measurement of $x_{t-s+1}$, subject to residual noise,
so that we have formal identification and a quantitative anchor for prior specification.

Beyond the need for priors for model hyper-parameters, 
we need structures for the  error volatility processes  $\sigma_{it}$ in
\eqns{DFM}{DTRFMj} and the  TVAR innovations variance process $w_t$ in \eqn{xt}.   
For routine analysis that is not inherently focused on volatility prediction,  standard Bayesian variance discount learning models-- 
effective random walks whose variability is controlled by a single discount factor-- 
are defaults. Specified to describe typically slowly, randomly changing variances, the  inverse gamma/beta Bayesian model  has the
ability to track   time-varying variances over time, and to deliver full posterior samples 
from relevant conditional posteriors for volatility sequences in MCMC analyses. We use variance discount models here, based on
standard theory in, for example,~\citet[][chap. 10]{WestHarrison97} and~\citet[][chap. 4]{PradoWest10}; these are simply specified via 
two discount factor hyper-parameters: $\lambda_\sigma$, for each of the set of 
observation volatilities, and  $\lambda_w$ for the TVAR innovations volatility. 

\subsection{Latent Components and Dynamic Transfer Responses \label{subsec:DTRFMcomps} }

Substantive  interpretation is aided by 
 investigating the more detailed structure that theoretically underlies the 
latent TVAR process $x_t.$  Specifically,  well-known (and well-exploited) time series decomposition 
theory~\citep[e.g][]{West1997,WestHarrison97,PradoWest10,West2013} shows that, given the model parameters, the
$x_t$ series has the decomposition 
	\begin{eqnarray}
	x_t = \sum_{g=1:\tilde p_t} \tilde x_{gt} + \sum_{h=1:\hat p_t} \hat x_{ht},  \label{eq:TVARdecomp} 
	\end{eqnarray}
where  the $\tilde x_{gt}, \ \hat x_{ht}$ are \lq\lq simpler'' component time series processes and $\tilde p_t, \hat p_t$
are non-negative integers such that $2\tilde p_t+\hat p_t=p.$  The values of these integers and the nature of the
component processes depend on the model parameters $\bdelta_t$. Typically, slow variation over time 
in these yields stable numbers  $\tilde p_t, \hat p_t$ and the resulting processes are computable directly from 
(posterior samples or estimates of) the $x_t$ and $\bdelta_t.$    The component processes $\tilde x_{gt}$ 
have the (approximate) forms of time-varying autoregressive moving averages--TVARMA(2,1) processes--
exhibiting quasi-periodic behavior:   each $\tilde x_{gt}$ is a stochastic sine wave whose 
amplitude, phase and frequency varies in time;   the time variation in frequency is directly related to that in $\bdelta_t$, 
while the amplitude and phase variation is inherent and driven by the levels of variation controlled by $w_t.$  Further, 
posterior inferences for the time-varying frequencies, amplitude and phase are directly available from posterior 
simulations that generate samples of the $x_t$ and $\bdelta_t$   at each time. 
In parallel,  each $\hat x_{ht}$ is a TVAR(1) process, with time variation in short-term autocorrelations 
driven by that in $\bdelta_t$. As with the $\tilde x_{jt},$ we have direct access to posterior inferences on the 
TVAR(1) parameters of these component processes from simulations of the posterior for $x_t,\bdelta_t$ at each time.  
This decomposition therefore gives inferences on underlying time-frequency 
and short-term dependency structures underlying $x_t$  and its dynamic behavior. 

From \eqns{DTRFMj}{TVARdecomp} it follows that 
$$
	y_{it} = \sum_{g=1:\tilde p_t} \tilde y_{igt} + \sum_{h=1:\hat p_t} \hat y_{iht} + \nu_{it}
$$
where, for each $g,h$ in the ranges displayed, 
$$
 \tilde y_{igt} = \sum_{k=1:r-1} \beta_{ikt} \tilde x_{gt}\qquad\textrm{and}\qquad  \tilde y_{iht} = \sum_{k=1:r-1} \beta_{ikt} \tilde x_{ht}.
$$
Thus the transfer response pattern defined by the time-varying factor loadings translates the nature of the 
inherent, underlying components of the \lq\lq driving'' $x_t$ process to each of the output/response variables. 

The above shows that this   class of models provides   broad scope for capturing 
multiple time-varying patterns of component structure-- including several or many components with 
dynamically varying time-frequency characteristics-- via a single latent process filtered to construct the
latent factor vector process $\f_t$ in the general framework.   The flexibility of these models for increasingly
high-dimensional response series $\y_t$ is then further enhanced through the ability of models with
series-specific and time-varying loadings $b_{ikt}$ to differentiate both instantaneous and time-varying 
patterns in the transfer responses.

\subsection{Model M+: DTRFM with a Time-Varying VAR Component  \label{subsec:TVVARDTRFM}}

A direct model extension adds back a non-zero dynamic regression term to provide an example of 
time-varying FAVAR models.   That is, with the dynamic factor component as specified via Model M, 
suppose $\y_t$ now follows \eqn{DFM} where $q=m,$ the $m\times m$ matrix $\A_t$ contains time-varying autoregressive parameters 
and $\z_t=\y_{t-1}.$   That is, $\y_t$ is dynamically regressed on the immediate past value $\y_{t-1}$ as 
well as the underlying components of a driving latent process through the dynamic transfer response mechanism: 
we denote this as a TV-VAR(1) component of the model. 

This extension of Model M allows for the transfer response effects of the fundamental, driving $x_t$ process to
be overlaid with spill-over effects between individual response series 
from one time point to the next, modeled by a basic TV-VAR(1) component. 
This can be regarded as a model extension  to assess whether the empirical TV-VAR component is able to
explain structure in the response data not adequately captured by the structure dynamic factor component.  For 
increasingly large $m,$ the TV-VAR(1) model component alone (i.e., setting $\B_t=\zero),$ 
implies what can be quite flexible marginal processes for the individual $y_{it};$ in contrast, the dynamic transfer response
factor component-- while also quite flexible-- represents structurally related processes. There is thus opportunity to
for evaluation of the latter in the extended Model M+.

\section{Dynamic Sparsity and Latent Thresholding \label{sec:LTMs}}

\subsection{Thresholding of Dynamic Parameter Processes} 

As the dimension $m$ of response variables and the number $r$ of effective latent factors increases, it
 becomes increasingly untenable to entertain models in which all loadings in $\B_t$ are non-zero.  Further, 
depending on context, it is also scientifically reasonable to entertain models in which one or more variables
may relate-- in a time-varying manner-- to a particular element of the latent factor vector for some 
periods of time,  but that the relationships may be practically negligible at other epochs.    This is the 
concept of dynamic sparsity:  a particular $b_{ikt}$ may be non-zero over multiple, disjoint time periods, 
and adequately  modeled by a specified stochastic process model when non-zero, but effectively zero 
in terms of the effect of $f_{kt}$ on $y_{it},$ in other periods.  The same idea applies to dynamic regression and/or
autoregressive parameters in $\A_t.$   Analysis that permits this will allow for adaptation over time to
zero/non-zero periods as well as to inference on actual values when non-zero. This includes 
extreme cases when a $b_{ikt}$ may be inferred as effectively zero (or non-zero) over the full time period of
interest. 

Dynamic latent thresholding~\citep{NakajimaWest10,NakajimaWest11} addresses this question of {\em time-varying sparsity} in
some generality; this approach is now developed in our context of dynamic transfer response factor models.
We anchor the development on basic AR(1) process models for the free elements of the 
dynamic factor loadings matrix $\B_t,$  recalling that the first row of elements is constrained to fixed (0/1) values
as noted in Section~\ref{subsec:DTRFM}.  For $i=2:m,\ k=1:r,$ the $b_{ikt}$ are modeled via what we 
denote by the LT-AR(1) processes defined as follows: 
	\begin{eqnarray}
	\label{eq:b_it}
	b_{ikt} &=& \beta_{ikt} s_{ikt} \quad \mathrm{with} \quad
	s_{ikt} = I(|\beta_{ikt}| \ge d_{ik}), 
	\end{eqnarray}
where the latent process $\beta_{ikt}$ is AR(1) with  
	\begin{eqnarray}
	\label{eq:state_beta}
	\beta_{ikt} &=& \mu_{ik} + \phi_{ik}(\beta_{ik,t-1} - \mu_{ik}) + \eta_{ikt}, \quad
	\eta_{ikt} \sim N(0,v_{ik}^2),
	\end{eqnarray}
and where $|\phi_{ik}|<1.$  The processes are assumed independent over $i,k.$ 
The latent threshold structure allows each time-varying factor loading to be shrunk fully 
to zero when its absolute value falls below a threshold $d_{ik}$. This way, a factor  loads in explaining a 
response only when the corresponding $\beta_{ikt}$ is ``large enough''. Inference on the latent  $\beta_{ikt}$ processes and
threshold parameters  make this data-adaptive, neatly embodying and yielding 
data-informed time-varying sparsity/shrinkage and parameter reduction.  

The same approach applies to the time-varying autoregressive parameters in the extension to Model M+. That is, 
the effective model parameters $\A_t = ( a_{ijt} )_{i,j=1:m}$ are modeled as thresholded values of AR(1) processes
$\alpha_{ijt}$ in precisely the same way as for the $b_{ikt}.$ Details are left to the reader as they follow the development
for $\B_t$ with simple notational changes. 

\subsection{Structured Priors on Thresholds} 

It will be evident that prior specification for threshold parameters $d_{ik}$ are key in defining
practical models.   We can do this by referencing the expected range of variation of the
corresponding $\beta_{ikt}$ process.  Under the AR(1) process model detailed above, 
$\beta_{ikt}$ has a stationary normal distribution with mean $\mu_{ik}$ and variance $u_{ik}^2 = v_{ik}^2/(1-\phi_{ik}^2).$ 
Given the hyper-parameters $(\mu_{ik},\phi_{ik},v_{ik}),$ this allows us to compute the probability that 
$\beta_{ikt}$ exceeds the threshold-- i.e., the probability of a practically significant coefficient-- across any range of possible
thresholds.   \cite{NakajimaWest10} follow this logic to specify informative, structured priors for $d_{ik}$ that 
depend explicitly on $(\mu_{ik},\phi_{ik},v_{ik}).$  We use this specification here; in particular, take
conditional uniform priors 
$$d_{ik} | K, \mu_{ik},\phi_{ik},v_{ik}   \sim U(0, |\mu_{ik}| + K u_i )$$ for some $K>0.$  
Direct evaluation then yields marginal (with respect to $d_{ik}$) sparsity probabilities
\begin{align*}
Pr(s_{ikt}=1|K) & \equiv Pr(|\beta_{ikt}|> d_{ik}|K) \\
                       &= 2-2 \Phi(K) - 2K^{-1} \phi(K) + K^{-1}2^{1/2}\pi^{-1/2}
\end{align*}
 where 
$\Phi(\cdot)$ is the standard normal cdf.  This is trivially evaluated. For large $K$, this is also 
very well approximated by $ K^{-1}2^{1/2}\pi^{-1/2}$ (this is  extremely 
accurate for $K$ as low as 2 and practically relevant values of $K$ exceeding that). 
The sparsity probability  strictly decreases in $K$ and decays to values of
about 0.25, 0.20 and 0.15, respectively,  at about $K=3.2, 4$ and 5.3, respectively.     
This gives us    assessment of what a particular choice of $K$ implies in terms
of overall expected levels of sparsity {\em a priori}.  In our studies, we find strong robustness in 
posterior inferences  to specified values of
$K$ above 3 or so, and use that value as a default.  Note also that there is flexibility to customize the prior to 
use different $K$ values for each threshold, to cater for contexts in which we aim to favor   higher thresholds (and
hence higher probabilities of zero parameter process values) for some $i,k$ than for others.

\subsection{MCMC-based Computation \label{sec:computation}} 

MCMC computations extend and modify those developed for time-varying autoregressions and multivariate stochastic volatility factor
models in~\cite{NakajimaWest10,NakajimaWest11}. The overall MCMC integrates a series of steps that use standard simulation 
components from Bayesian state space models~\citep[e.g.,][]{WestHarrison97, PradoWest10} and from traditional (static loadings) latent factor models~\citep{AguilarWest00,LopesWest04}. Customization here involves modifications to resample the latent TVAR factor process
in our dynamic transfer responses factor context, and other elements including 
Metropolis Hastings steps as in~\cite{NakajimaWest10} for the latent threshold components.   
The Appendix accompanying this paper describes key details, and notes how the MCMC directly extends previously described strategies for
dynamic latent threshold models.

\section{Application:  EEG Time Series Analysis   \label{sec:EEG}}

\subsection{Background, Data and Prior Modeling Approaches  \label{sec:EEGintro}}

Electroencephalographic (EEG) traces are time series of electrical potential fluctuations at various scalp locations of a 
human subject, reflecting the complex dynamics of underlying neural communication. Analysis of multichannel EEG traces is key to  
understanding the impact of electroconvulsive therapy (ECT), one of the most effective treatments known for major depression  
with electrically induced seizures in patients~\citep[][]{WeinerKrystal94}. The convulsive seizure activity drives multichannel EEG traces 
and statistical interest is to model such multivariate time series   in order to reveal underlying characteristics and effects of ECT. Various  
models have been studied to explore features of EEG time 
series~\citep[e.g.,][]{KitagawaGersch96, WestPradoKrystal99, PradoWestKrystal01, Prado10a, Prado10b, PradoWest10}. 
Univariate TVAR models are well-used and proven as models of individual EEG channels~\citep[e.g.][]{WestPradoKrystal99, PradoWestKrystal01}; they can adequately represent what can be considerable changes in the patterns of evolution of time-frequency structure in
such series, as well as differences and changes in relationships across the EEG channels.  Such studies highlight the need for
multivariate models of the time-varying commonalities across channels, with latent process structure reflecting the inherent, 
underlying mechanisms of neural communication. 

Our analysis adapts the earlier  approach of~\cite{PradoWestKrystal01}. 
That work was the first Bayesian approach to multivariate time series analysis that incorporated the key scientific 
feature of a single, focal latent process \lq\lq driving'' the EEG signals across multiple channels. The authors used 
a novel dynamic distributed lag approach that aimed to capture time-varying lag-lead structures across the EEG channels,
introducing a customized model specific to that context. Though effective, that approach was very specific and empirical-- the
authors developed dynamic regressions of $m-1$ of the channels on the {\em  observed signal} of one selected channel, the
latter chosen as an empirical proxy for the underlying latent driving process $x_t.$  The developments of the current paper 
provide a general, flexible and-- in terms of the specific goals of the dynamic lag/lead study-- almost perfectly suited context that can be seen, in part, 
as an outgrowth from that prior work. Here the identification of dynamically adaptive lag/lead structure is driven by combining 
time variation in non-zero factor loadings with the latent threshold approach.

\begin{figure}[htpb!]

\centering
\includegraphics[width=2in]{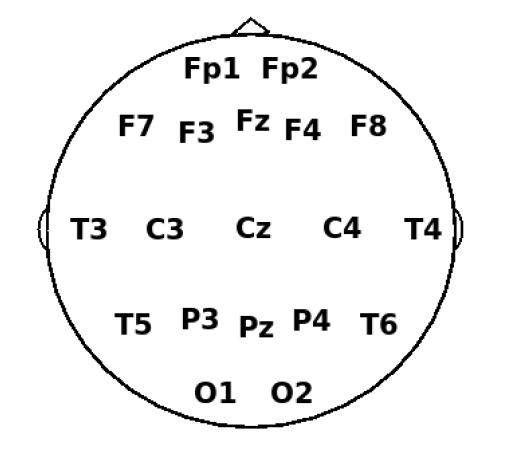}
\caption{Representation of the 19-electrodes placement over the scalp. The $m=19$ series are 
measurements of electrical potential fluctuations taken in parallel at each of these locations, defining the EEG channels (International 10-20 EEG System).}
\label{fig:map}
\end{figure}

\begin{figure}[tph!]

\centering
\includegraphics[width=5in]{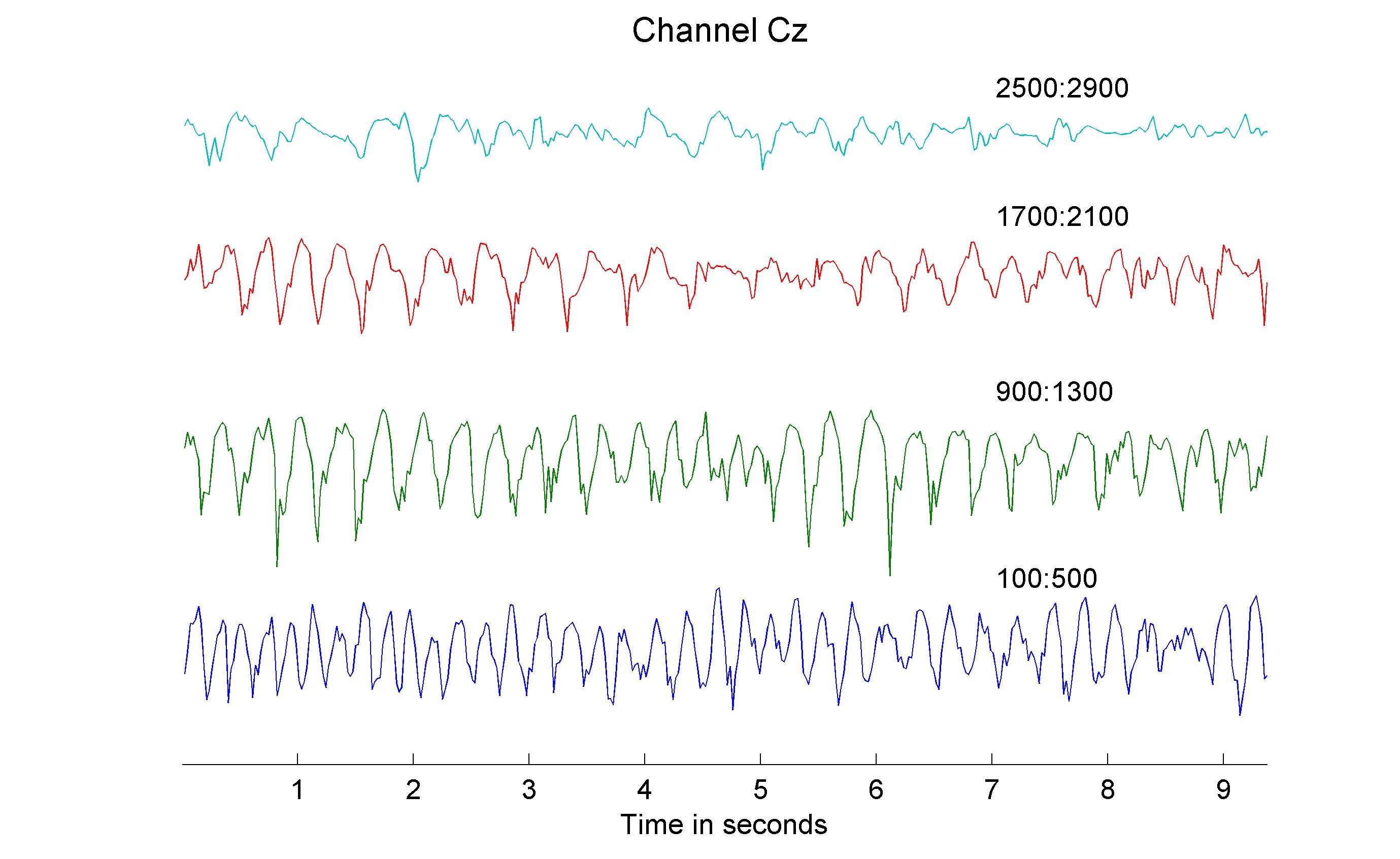}\\
\includegraphics[width=5in]{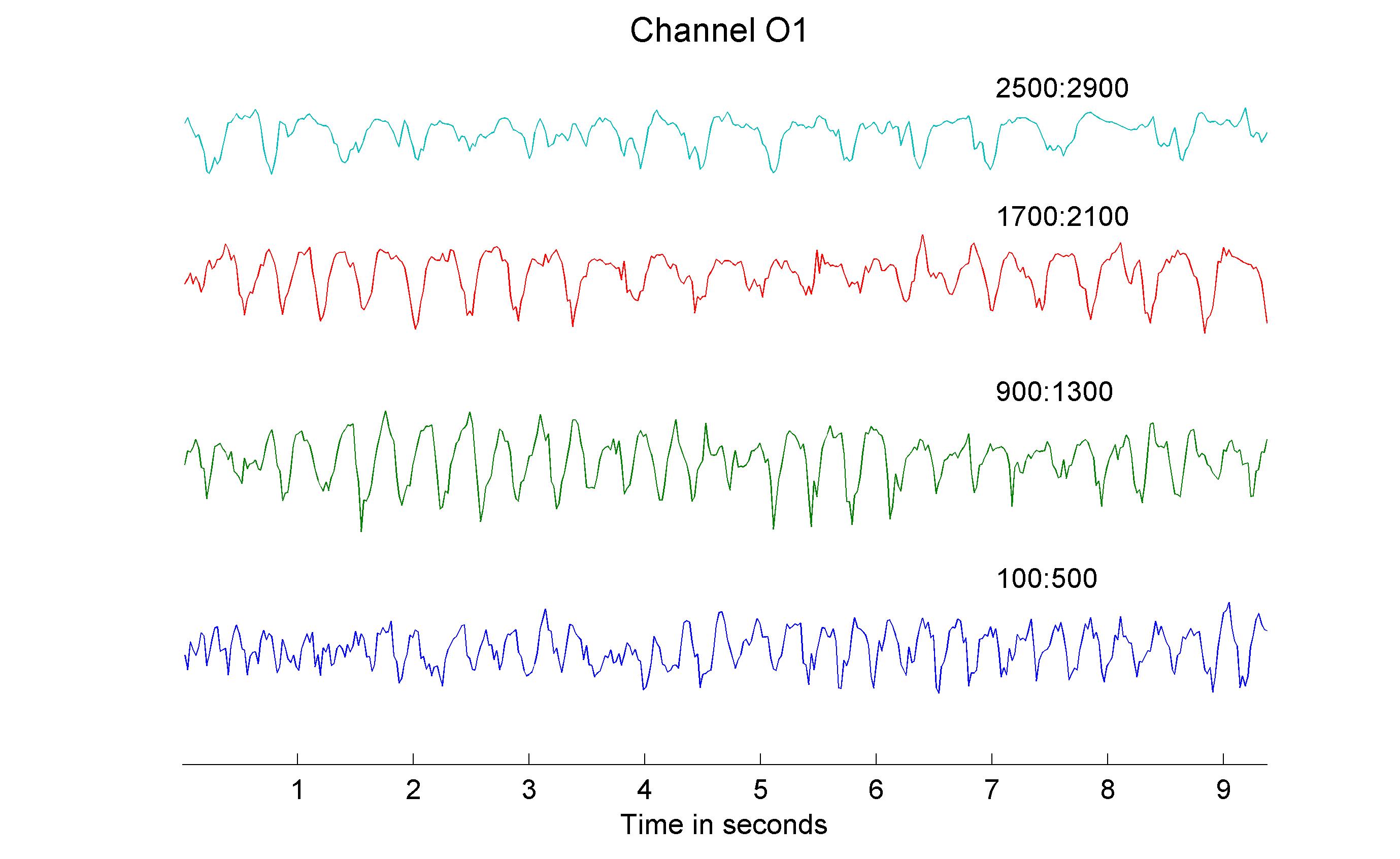}
\caption{Sections of time series data on two selected EEG channels, with standardization so that the
vertical scales of variation in EEG electrical potential are comparable across epochs and channels.}
\label{fig:EEGdata}
\end{figure}

The study here explores $m=19$-channel EEG times series recorded in one seizure of one patient, as reported and 
earlier analyzed in~\cite{WestPradoKrystal99} and \cite{PradoWestKrystal01}. 
The EEG channels are $m=19$ electrodes located around and over the patient's scalp; see Figure~\ref{fig:map}. 
The original data set has sampling rate 256Hz over a period of 1-2 minutes;  following and to compare directly with~\cite{PradoWestKrystal01}, 
we analyze the series subsampled every sixth observation after removing about 2{,}000 observations from the beginning (up to a higher amplitude portion of the seizure)
yielding $T=3{,}000$ observations.   Representative graphs of data on two of the channels over selected epochs appear in Figure~\ref{fig:EEGdata}.  
Visual inspection of the substantial time-varying, quasi-periodic trajectories of the data 
indicates that signals on some EEG channels are obviously \lq\lq delayed'' with respect to other channels, and the apparent delays (lags) 
vary substantially through time. This is perfectly consistent with the dynamic patterns of relationships of individual channels (the $y_{it})$ with 
an underlying seizure process (the latent $x_t$) captured by our model structure (Section~\ref{subsec:DTRFMcomps});  
 the latent $x_t$ process  represents a range of dynamic quasi-periodicities characterizing multiple
brain wave components overlaid by, and modified by, the induced seizure, and the time-varying lag/lead relationships among 
channels are represented by channel-specific and time-varying factor loadings, some of which may be negligible for all time or
for periods of time, and relevant elsewhere. 

\subsection{DTRFM- Model M for Multivariate EEG Signals  \label{sec:EEG-DTRFM}}

Our analysis summaries are based on $r=5$ effective lags, i.e., the model has a $5-$dimensional 
latent factor vector $\f_t'=(x_t,x_{t-1},\ldots,x_{t-4})'$ and the first row of $\B_t$ set to $(0,0,1,0,0)$ 
 as the basis for model identification.  This precisely parallels the setup in the
empirical model of~\cite{PradoWestKrystal01}.  As discussed in Section~\ref{subsec:DTRFM}, 
some constraints of this form are needed on elements of $\B_t$ to  formally identify the single latent factor process 
model.  There is no  loss of generality nor any superfluous structure imposed on the model here;  
we could choose any element of the first row of $\B_t$ to insert the 1, with different choices simply shifting the 
implied time origin of the $x_t$ process. Under this structure, the first EEG channel $y_{1t}$ loads only $x_{t-2}$, while the
other channels have loadings in the first (last) two columns of $\B_t$ related to the leading (lagged) values of the $x_t$ process.

Our analysis takes the so-called vertex channel Cz as series $i=1$.  See Figure~\ref{fig:map}. This parallels the use of the observed data on 
this specific channel as an empirical factor process in~\cite{PradoWestKrystal01}.   The other channels are ordered from
the centre out.  One further modeling detail relates to a modification for a further, subtler \lq\lq soft'' identification question.  The model so 
far implies that $y_{1t} = x_{t-2}+\nu_{1t},$ so the conditional variation expected in channel 1 is the sum of
time-varying contributions from the  $x_t$ process plus $\sigma_{1t}^2.$ As in all state-space models with multiple
components contributing to variability in observed data, distinguishing and partitioning the contributions requires care in prior
specification; the picture is complicated here as time variation in $\sigma_{1t}^2$ \lq\lq competes''  with the intricate
dynamics of the $\bdelta_t$ and time variation in $w_t$.  A specification that controls this more usefully in the current latent 
factor models is to fix as constant the
measurement error in series 1, i.e., set $\sigma_{1t}=\sigma_1,$ constant over time.  This ensures the interpretation of 
$\nu_{1t}$ as pure measurement error (there being no reason to expect time variation in pure measurement error variances, as opposed to the characteristics of  the underlying factor processes and transfer response/loadings parameters). We do this, maintaining the stochastic variance discount model for the other $\sigma_{jt}, $ $(j=2:m);$
the latter combine pure measurement error and any other identified changes in residual volatility across these channels. Then, 
posterior inferences indicating substantial patterns of time variation in the latter then indicate the ability of the discount models to 
account for relative variability not captured by the underlying, identified latent factor process. The MCMC analysis of Section~\ref{sec:computation}
is trivially modified; a traditional inverse gamma prior on $\sigma_1^{2}$ leads to an inverse gamma  complete conditional posterior.  

The analyses summarized are based on model order $p=6$ for the latent $x_t$ process. While formal order selection approaches 
could be entertained~\citep[e.g.,][]{HuertaWest99, PradoHuerta02}, our philosophy based on applied work with TVAR and related models in 
many areas is to fit successively larger models and assess practical relevant of resulting inferences. Here we fit the DTRFM 
with model orders up to $p=12,$ and for each analysis examine the posterior estimates of components $\tilde x_{jt}, \ \hat x_{jt}$
as detailed in Section~\ref{subsec:DTRFMcomps}.  With successively higher values of model order $p,$ we find robust estimation 
of $\tilde p_t=3$ quasi-periodic components with estimated frequencies varying over time in ranges consistent with known ranges of
seizure  and  normal brain wave activity.  Model order $p=6$ is needed to identify these three components, and they
persist in models of higher order; in addition to their substantive relevant, the estimated components are sustained over time and
vary at practically relevant levels in terms of their contributions to each of the EEG series.  However, moving beyond $p=6$ leads to 
increasing numbers of estimated components that are very ephemeral, of  low amplitudes and higher inferred frequencies beyond the range of substantive relevance. This signals over-fitting as the model caters to finer aspects of what is really noise in the data. 

and then disregarding such estimated noise components  is certainly acceptable,  we prefer to cut-back to
the model order $p=6$ that identifies  the main component structures without these practically \lq\lq spurious'' elements. 

Model specification is completed with priors for hyper-parameters.      
We take   $\sigma_1^{-2} \sim G(500, 10^4)$, supporting a range of values for 
$\sigma_1$ and with prior mean  for $\sigma_1$ near 4.5.  Seizure EEG data typically range over 300-600 units on the potential scale, with sample 
standard deviations over selected epochs varying from 40-100 or more.  Hence an expectation of measurement 
error standard deviation around 4-5 is consistent with prior expectations that measurement error constitutes in the range of 4-12\% of the 
signal uncertainty in the traces.
For the stochastic discount variance models, we set values of the discount factors as  
$\lambda_w = \lambda_\sigma = 0.99$; this is based in part on examination of analyses with various 
 values, and consistent with relatively modest levels of volatility in variance components.
Priors for the hyper-parameters of the latent AR(1) parameter  processes are as follows:  
$1/v_{ik}^2 \sim G(50, 0.01)$,  $(\phi_{ik} + 1)/2 \sim B(20, 1.5)$, and $\mu_{ik} \sim N(0, 1),$ independently, for $i=2:m,\ k=1:r$. 
This anticipates persistence in non-thresholded latent factor loadings, while allowing for some of the
loadings to exhibit notable patterns of change.   Finally, we take $\bPsi^{-1} \sim W(100, 10^{-3}I)$ and
set $K=3$ in the conditional uniform priors for thresholds.

\subsection{Some Posterior Summaries from Analysis of DTRFM- Model M\label{sec:EEG-M}}

Summaries here come from $J=20{,}000$ MCMC draws after a burn-in period of $5{,}000$. 
Computations were performed using custom code in Ox~\citep{Doornik06}.

\begin{figure}[htpb!]

\centering
\includegraphics[width=5in]{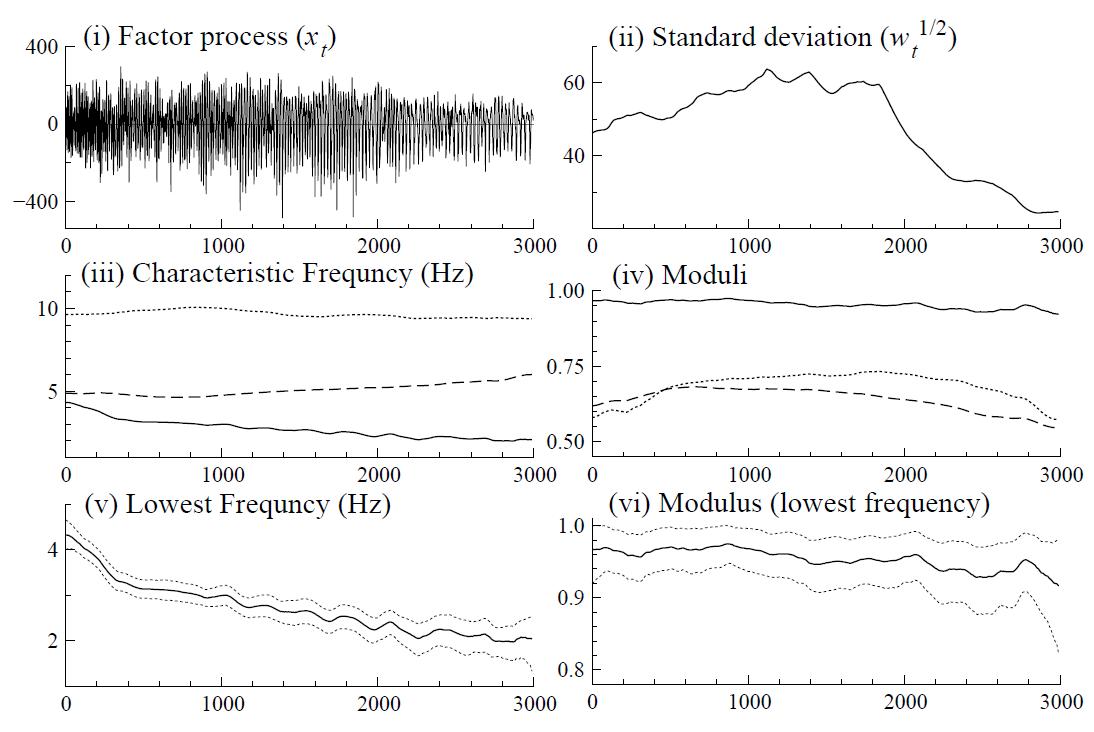}

\caption{Elements of inference on the latent factor process and its components in the EEG analysis. The trajectories are 
posterior means of:  (i) the factor process $x_t,$ and (ii) the innovation volatility 
process $w_t^{1/2}$; (iii)  the characteristic frequencies, and (iv) moduli of the three quasi-periodic components $\tilde x_{jt},$ $j=1:3;$ and
 (v) the characteristic frequency, and (vi) modulus of the lowest-frequency component (solid) with 95\% credible intervals (dotted).}
\label{fig:factor}
\centering
\includegraphics[width=5in]{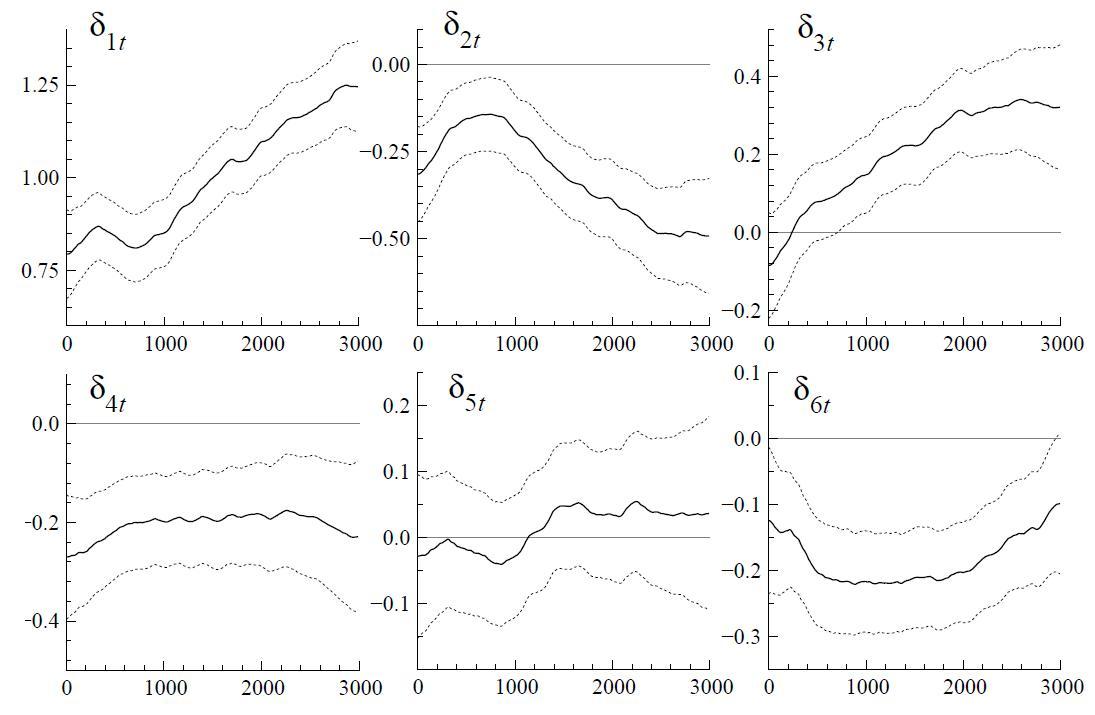}
\caption{EEG analysis: Trajectories of posterior means (solid) and 95\% credible intervals (dotted) for the TVAR coefficients $\delta_{1:6,t}$. }
\label{fig:tvar}
\end{figure}

Figure \ref{fig:factor} displays time trajectories of the posterior means of the factor process $x_t$, and the volatility 
$w_t^{1/2}$ of its driving innovations.  The figure displays similar trajectories for 
the time-varying characteristic frequency and modulus for each of the three identified  quasi-periodic components in $x_t$, the 
$\tilde x_{jt}$ of Section~\ref{subsec:DTRFMcomps}. 
The $\tilde x_{jt}$ component of lowest frequency has oscillations in the so-called \lq\lq seizure'' or 
\lq\lq slow wave'' band,   considerably decaying toward the end of the seizure episode. Notably, the other two inferred
components have characteristic frequencies and moduli that are rather stable over time though exhibit minor variation.

The frequency trajectories of the three quasi-periodic components show that each  lies in one of the expected  
neuropsychiatric categories: the so-called {\it delta} band (roughly 0-4Hz), {\it theta} band (4-8Hz), and {\it alpha} band (8-13Hz)~\citep{Dyro89}. 
Each component process $\tilde x_{jt}$ is defined by the corresponding characteristic frequency, while being  broad-band in the spectral domain with time-varying spectra that can be understood to peak at the characteristic 
frequencies themselves. 
The lowest-frequency component stays in the {\it delta} range and gradually decreases over time; its modulus is close to one (solid line in Figure \ref{fig:factor}(iv)), which indicates a considerably persistent component; this so-called delta-slow wave dominates the factor process during the course of the seizure, while its frequency content slides towards lower values towards the end of the seizure episode.
The other two quasi-periodic components lie in the {\it theta} and {\it alpha} ranges; 
their moduli and amplitudes are  lower than those of the dominant 
component over the whole seizure course, and
show only limited changes over time. These
reflect known frequency ranges of normal brain signaling, being dominated through much of the period
by the strong seizure waveform. The innovations volatility rises in the initial part of the seizure to drive increased amplitude 
fluctuations throughout the central section, and then decays in later stages corresponding to the control and 
dissipation of the brain seizure effects. 
These features are consistent with 
expected structure in the seizure process, and with the broad results of~\cite{PradoWestKrystal01}.

Figure \ref{fig:tvar} provides the trajectories of the posterior means and 95\% credible intervals for the TVAR 
coefficients $\delta_{1:p,t}$. All are markedly time-varying.  
The 95\% credible intervals are slightly wider during the late time periods, which feeds through to increased
 uncertainties in features of the quasi-periodic components; Figures \ref{fig:factor}(v) and (vi), displaying
  the posterior means and 95\% credible intervals of the frequency and modulus for the 
  lowest-frequency component respectively, showing somewhat 
increased uncertainties towards  the end of the seizure.
\begin{figure}[hb!]

\centering
\includegraphics[width=5in]{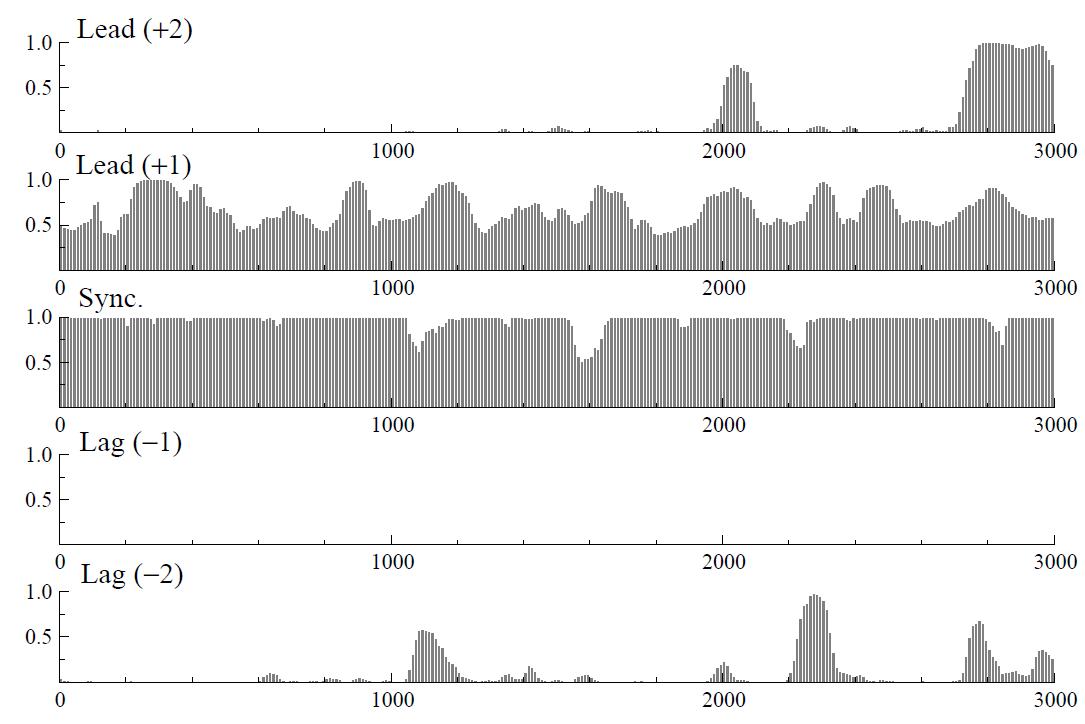}

\caption{EEG analysis: Trajectories of posterior probabilities $Pr(s_{ikt}=1|\y_{1:T})$ for channel F7,  
index $i=4$,  indicating the inferred probabilities of lag-lead structure in transfer response to 
$x_t$ as they vary over time.}
\label{fig:s}
\end{figure}

To generate some insights into the nature of dynamic sparsity under the latent threshold model,  we select one channel-- F7 at $i=4$,-- and plot 
the corresponding trajectories of the estimated posterior probabilities $Pr(s_{ikt}=1|\y_{1:T})$ over time; i.e., the probability of a non-zero 
loading of channel F7 on each of the values $x_{t-k}$ for $k=0:r-1.$ See 
Figure~\ref{fig:s} where we indicate the loadings on $x_t, x_{t-1}$ by Lead($+2$) and ($+1$) respectively, 
that on $x_{t-2}$ by Sync, and those on $x_{t-3}, x_{t-4}$ by Lag($-1$) and ($-2$) respectively. The annotation 
here refers to lead/lag relative to the vertex location Cz that reads-out an unbiased estimate of $x_{t-2}$.  So
a non-zero loading of F7 on $x_t$, for example, defines a 2-period lead of that channel relative to the vertex,
whereas a non-zero loading on $x_{t-3}$ represents a 1-period lag relative to the vertex channel Cz, and so forth. 
From the figure, it is clearly inferred that there is strong synchrony  between F7 and Cz in their transfer responses
to fluctuations in $x_t$ based on the Sync trajectory. Also, F7 also has a reasonable probability of responding to 
the latent factor process $x_t$ 1-period ahead of Cz, and 
almost surely does not lag Cz in the transfer response over most of the time period, nor lead by more than 1 period
until perhaps later in the seizure episode. The ability of latent thresholding to adaptively indicate existence of 
non-zero loadings during some periods and not others, while also operating as a \lq\lq global'' variable selection 
mechanism as well, is nicely exemplified here.

\begin{figure}[hb!]

\centering
\includegraphics[width=5in]{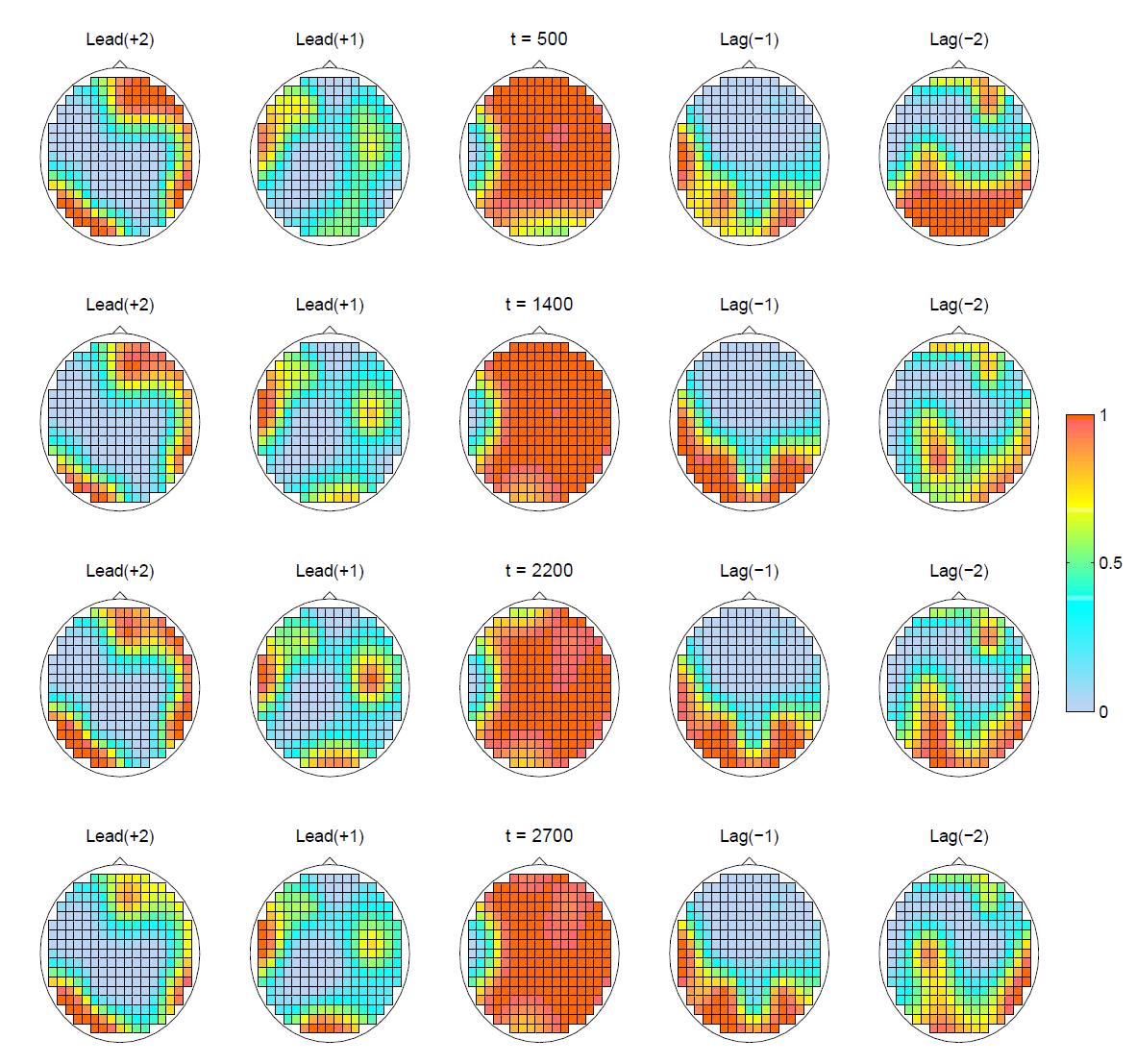}

\caption{EEG analysis: Contoured values of $Pr(s_{ikt}=1|\y_{1:T})$  interpolating from 
values at the channels $i=1:19.$  Each row corresponds to the values at a selected time point, and 
the columns represent the indices $k$  in relation to the transfer responses to $x_{t-k}$ for $k=0:r-1.$}
\label{fig:st}
\end{figure}

Figure~\ref{fig:st} provides a visual display of posterior probabilities $Pr(s_{ikt}=1|\y_{1:T})$ across all the channels $i=1:19$, and drawn at selected snapshots 
in time, with images created by linearly interpolating between the estimates at the electrode locations. Note that the model 
says nothing about spatial relationships between channels. The marked patterns of shrinkage in the latent threshold
model analysis does nevertheless indicate strong spatial relationships, while the relationships also show marked 
patterns of change over time.  For example, loadings of Lead(+2) are commonly and globally shrunk to zero from left frontal to right occipital sites. The Lead(+2) loadings around right frontal and prefrontal areas exhibit evolving degree of shrinkage. Similar changes are found in the parietal and occipital regions of Lag(-2) loadings. Meanwhile, almost no shrinkage is found in the synchronized loadings except for the channel T3 (left temporal).

\begin{figure}[p]

\centering
\includegraphics[width=5in]{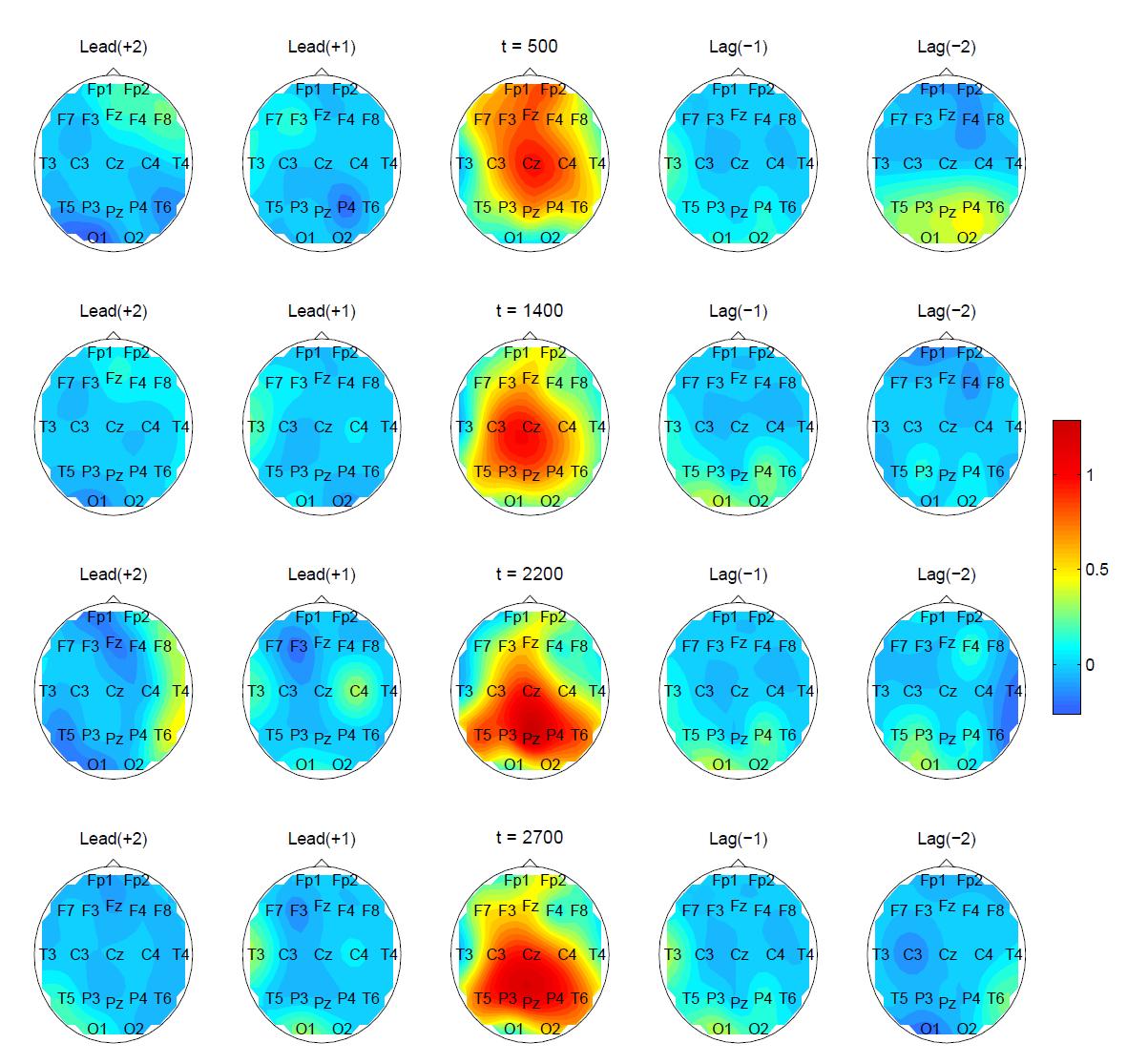}

\caption{EEG analysis: Estimated factor loadings $\hat b_{ikt}$  with the lag-lead structure at selected time points. The row-column layout of images 
corresponds to that in Figure~\ref{fig:st}. }
\label{fig:beta}
\end{figure}

Figure~\ref{fig:beta} is a companion  to Figure~\ref{fig:st} that exhibits aspects of  
estimated factor loadings with the lag-lead structure at selected time points.  The   images represent
estimates $\hat b_{ikt} = E(\beta_{ikt}| \y_{1:T}) \hat s_{ikt}$ where $\hat s_{ikt} =1$ if $Pr(s_{ikt}=1|\y_{1:T})>0.5 $ and zero otherwise.
Recall that the factor loading of  vertex channel Cz is fixed at 1 for the basis and 0 for lagged/leaded times. The estimates show strong patterns
of positive spatial dependencies with Cz at the synchronized state (zero lag/lead), with   concurrent loadings on the $x_t$ process
decaying towards the exterior regions. The approximate centroid of the higher loadings region moves from front to back through the course of the seizure,   consistent with what is understood to be the typical progression of seizure waveforms~\citep{PradoWestKrystal01}.
In the third row of the figure ($t=2{,}200$), the highest loadings appears at and near channel Pz, and the parietal region exhibits rising intensity. Another higher intensity is detected around the right temporal area in Lead(+2) and in the channel C4 in Lead(+1). This indicates dynamics of the driving latent process exhibited earlier in right temporal/central areas and followed  in the occipital region; 
this spatial asymmetry in estimated transfer response loadings again links to experimental expectations for the progression of 
seizure activity. In the last row of the figure ($t=2{,}700$, a late stage of the seizure), the major lead/lag loadings diminish 
while the synchronized loadings persist.

Two animated figures, available as online supplementary material,  
provide  more insight into the patterns of variation over time in factor loadings, the differences across
channels, and  the nature of the dynamic latent thresholding in particular.    
The first animation 
\href{http://www.stat.duke.edu/~mw/.downloads/NakajimaWest2016EEG/animate-st.avi}{(linked at the external site here)}
shows a movie of patterns of $Pr(s_{ikt}=1|\y_{1:T})$  interpolating from 
values at the channels $i=1:19.$  This shows how these patterns evolve over time $t$, providing a dynamic display 
from which the snapshots in Figure~\ref{fig:st} are selected at four specific times.  The second animation
\href{http://www.stat.duke.edu/~mw/.downloads/NakajimaWest2016EEG/animate-betat.avi}{(linked at the external site here)}
shows the corresponding movie for the interpolated estimates of factor loadings $\hat b_{ikt}$  
over all time; the  snapshots in Figure~\ref{fig:beta} are selected at four specific times. 
The animations clearly show and highlight regions of the brain surface where 
there is very low or negligible probability of lag or lead effects of the $x_t$ process,  other regions where sustained 
effects are very evident and regions in  which there is more uncertainty about potential effects,
together with inferences on the quantified lag/lead effects
in terms of the temporal evolution of the spatial patterns in estimated factor loadings.

\begin{figure}[htb]

\centering
\includegraphics[width=3.2in]{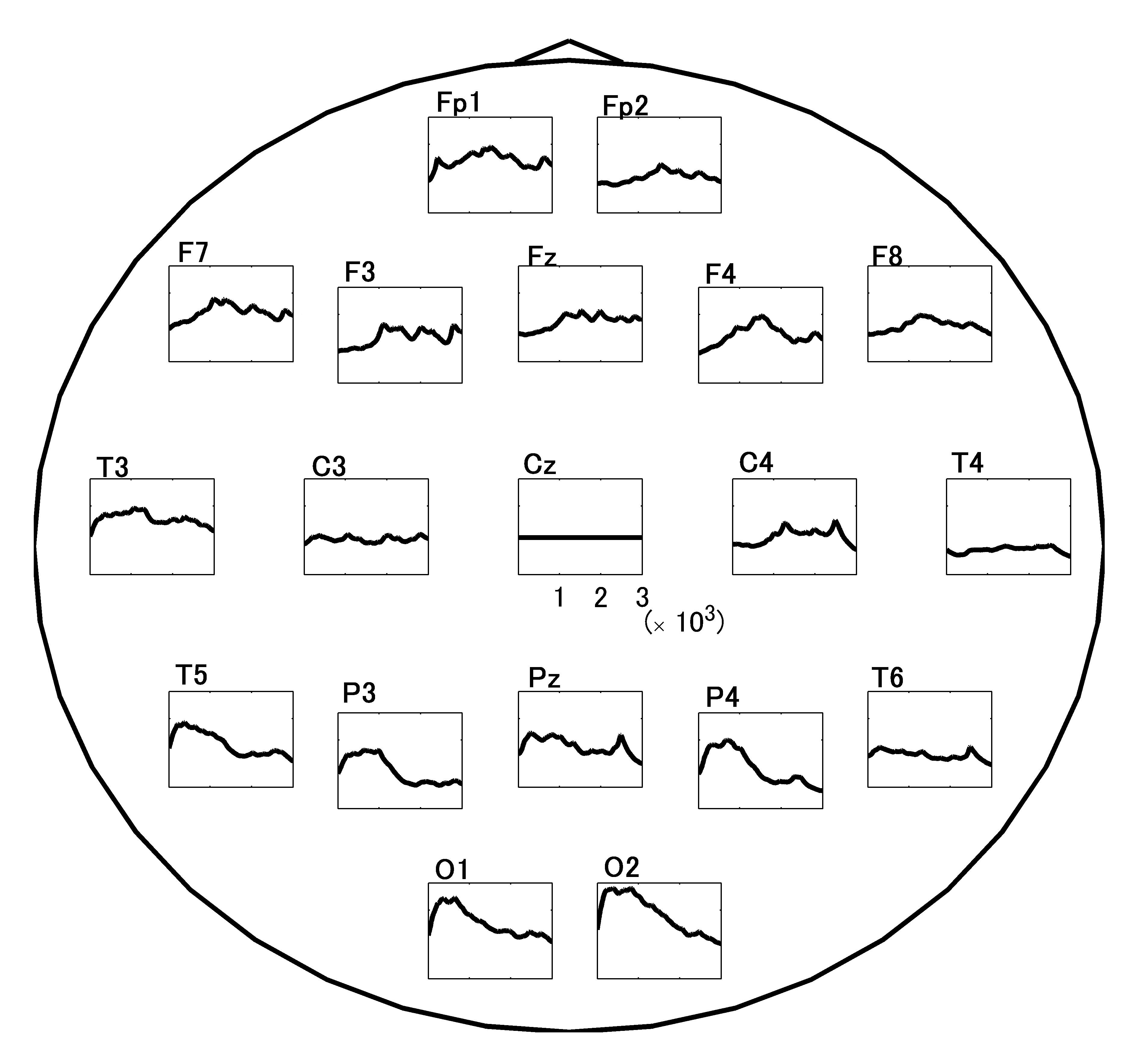}

\caption{EEG analysis: Trajectories of the posterior means of time-varying 
standard deviations of the idiosyncratic shocks, $\sigma_{ikt},$ 
across all channels $i=1:19.$ For clarity in presentation, the y-scale has been omitted; each standard deviation is graphed on the
scale of $0-50$ for comparability across channels.
Recall that the anchor channel Cz has constant standard deviation representing pure measurement 
error around the latent process at that channel.  Each graph is roughly located at the corresponding electrode placement
and the $x$-axes represent the full time period $t=1:3{,}000$.}
\label{fig:sigma}
\end{figure}

Figure~\ref{fig:sigma} plots $E(\sigma_{ikt}|\y_{1:T}), $ i.e., estimated standard deviations of the idiosyncratic shocks
in each channel  $i=1:19.$ Each graph is roughly located at the corresponding electrode placement. Recall that $\sigma_{1t}=\sigma_1$, the innovation standard deviation for the channel Cz, is assumed time-invariant, representing measurement error only, as part of
the model specification to define and identify the latent driving process $x_t$. The model 
allows for potential variations over time in standard deviations at other channels, with opportunity to identify
variability in the data not already captured through the time-varying loadings and latent process structure.

From Figure~\ref{fig:sigma}, there do appear to be variations across channels and they show some local spatial dependence. 
Trajectories of the neighboring channels F4 and F8 are clearly similar, exhibiting a major hike in the middle of the seizure. It is evident that some parietal and occipital sites (T3, P3, O1, O2 and P4) share a common trajectory, which marks a peak in an early stage of the seizure then gradually decrease towards the end of the seizure. As seen in Figures~\ref{fig:st} and \ref{fig:beta}, these sites also share some relationships in the latent threshold-induced shrinkage and loadings at Lag(-1). Further, the estimate shows similarities among the channels Pz, C4 and T6, whose patterns differ from those in the occipital region. This suggests an intrinsic difference between the central sites (Pz, C4 and T6) and the occipital sites (T3, P3, O1, O2 and P4), also suggested by 
Figures~\ref{fig:st} and \ref{fig:beta}. Across all but the vertex channel $Cz$ at $i=1,$ the idiosyncratic error terms 
$\nu_{it}$ represent a compound of measurement error and of additional patterns including local sub-activity of the seizure that is not explained by the latent factor process. There are also experimental and physiological noise sources that are likely to
induce local/spatial residual dependencies in the data not forming part of the main driving process $x_t,$ including 
electrical recording/power line noise and scalp-electrode impedance characteristics; these presumably 
also contribute to the time-variation patterns in the $\sigma_{it}$ identified and their spatial dependencies.

\subsection{Summaries from Analysis of Extended DTRFM- Model M+\label{sec:EEG-M+}}

Model M+ has 
	\begin{eqnarray}
	\y_t  = \A_t\y_{t-1} + \B_t \f_t + \bnu_t,     \qquad  \bnu_t  \sim N(\zero,\bSigma_t), \label{eq:LTFP_VAR}
	\end{eqnarray}
where $\A_t$ is the $m\times m$ matrix of lag-1 time-varying coefficients modeled using latent threshold AR(1) 
processes. Model M+ extends  Model M to potentially capture data structure not fully explained by the factor and residual 
component.  One interest is to more structurally explain the time variation in estimated residual volatilities $\sigma_{it}$ 
exhibited in the analysis of the baseline Model M.  A contextual question is that of representing potential \lq\lq spill-over'' 
 effects between EEG channels as the seizure waves cascade around the brain; that is, local (in terms of the 
neural network and perhaps in part, though not necessarily, physically spatially) transmission of signals between subsets of
channels that represent delayed responses to the latent $x_t$ process not already captured by the dynamic latent factor model
form.  The matrix $\A_t$ is expected to be sparse and modeled via latent threshold dynamic models, as earlier described.

\begin{figure}[hpt!]
\centering
\includegraphics[width=5in]{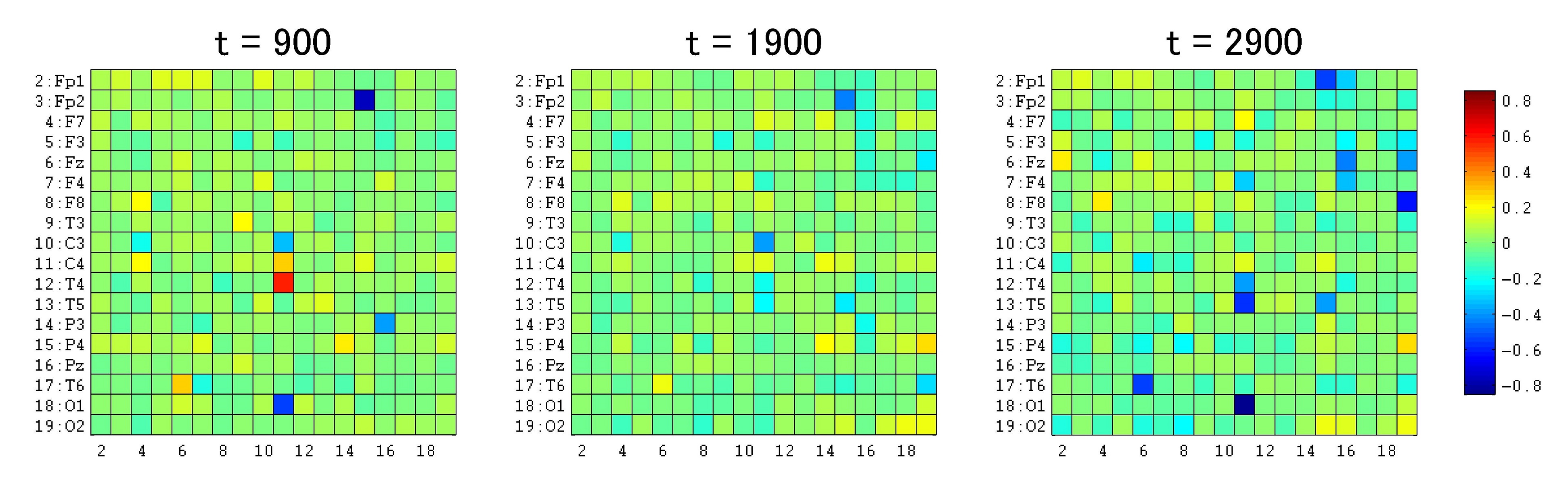} \\
\includegraphics[width=5in]{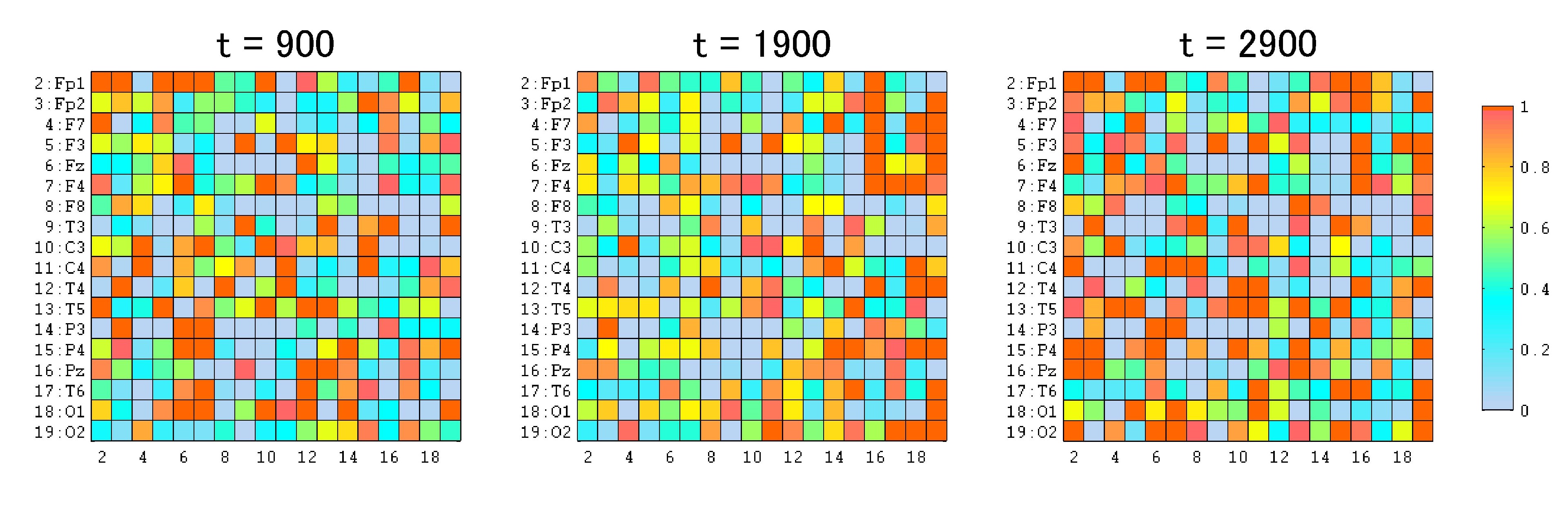}
\caption{EEG analyses: Posterior means of $\{\alpha_{ikt}\}$ (top) and posterior probabilities $Pr(s_{ikt}^a=1|\y_{1:T})$ (bottom) for $\A_t$, in the TV-VAR extended model M+.}
\label{fig:at}

\vskip.4in

\centering
\includegraphics[width=3.2in]{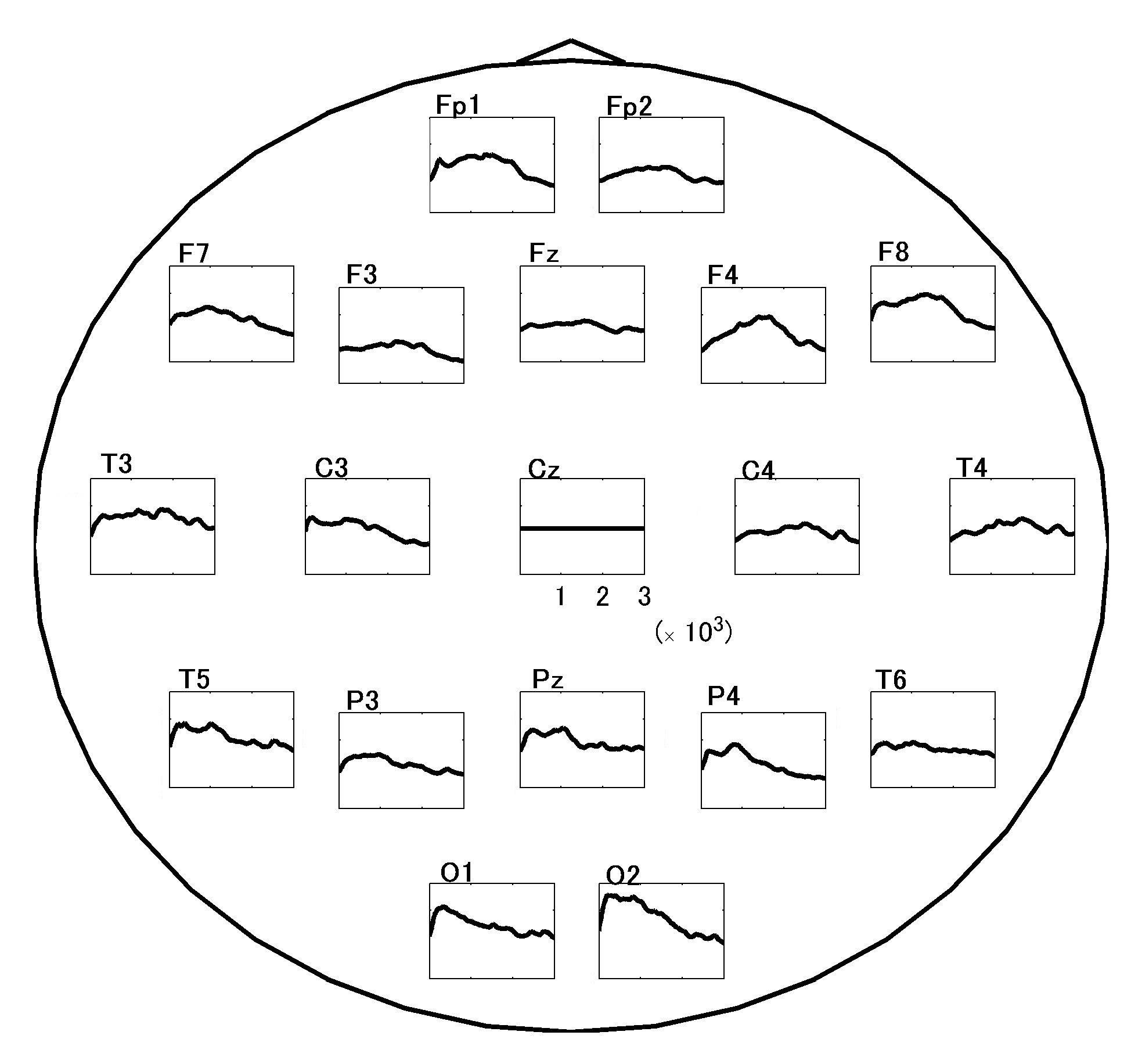}
\caption{EEG analysis: Trajectories of the posterior means of time-varying 
standard deviations of the idiosyncratic shocks, $\sigma_{ikt},$ 
across all channels $i=1:19$ for the extended Model M+. Details as in as in Figure~\ref{fig:sigma} for model M.}
\label{fig:sigma-var}
\end{figure}

Figure \ref{fig:at} plots the posterior means of the states $\{\alpha_{ikt}\}$ and the posterior probabilities $Pr(s_{ikt}^a=1|\y_{1:T})$, 
where $s_{ikt}^a = I(|\alpha_{ikt}| \ge d_{ik}^a)$ and $d_{ik}^a$ is the latent threshold for each state $\alpha_{ikt}$. 
The matrix $\A_t$ is evidently sparse and exhibits considerable changes in the state and the posterior shrinkage probability among the selected time points. Figure \ref{fig:sigma-var} shows the estimated standard deviations of the idiosyncratic shocks $E(\sigma_{ikt}|\y_{1:T})$. Compared with Figure \ref{fig:sigma}, the trajectories of standard deviations are generally somewhat smoother 
over time; some of the variation in the data not already captured by the $\B_t\f_t$ is now absorbed by the $\A_t\y_{t-1}$.

To explore some practical  implications of the extended Model M+ and compare with the baseline Model M, 
one aspect of interest is predicted behavior of the time series based on {\em impulse response analysis}
relative to the underlying $x_t$ process.  Standing at a current, specified time $t,$ this simply asks about 
the nature of expected development of the series $\y_{t+1:t+h}$ over the next $h$ time points based on an
assumed level of the \lq\lq impulse'' $\epsilon_{t+1}= e$ to the driving innovations of the latent process.  
In applied work in economics,  impulse responses are often primary vehicles for communicating 
model implications, comparing models, and feeding into decisions.   The use of latent thresholding in 
 macroeconomic models has focused on this, in part, and  clearly demonstrated the utility
of dynamic latent thresholding in inducing more accurate predictions and, in particular, more statistically and
substantively reliable impulse response analyses~\citep{NakajimaWest10}.

\begin{figure}[p]

\centering
(i) Model M \\
\includegraphics[width=3.8in]{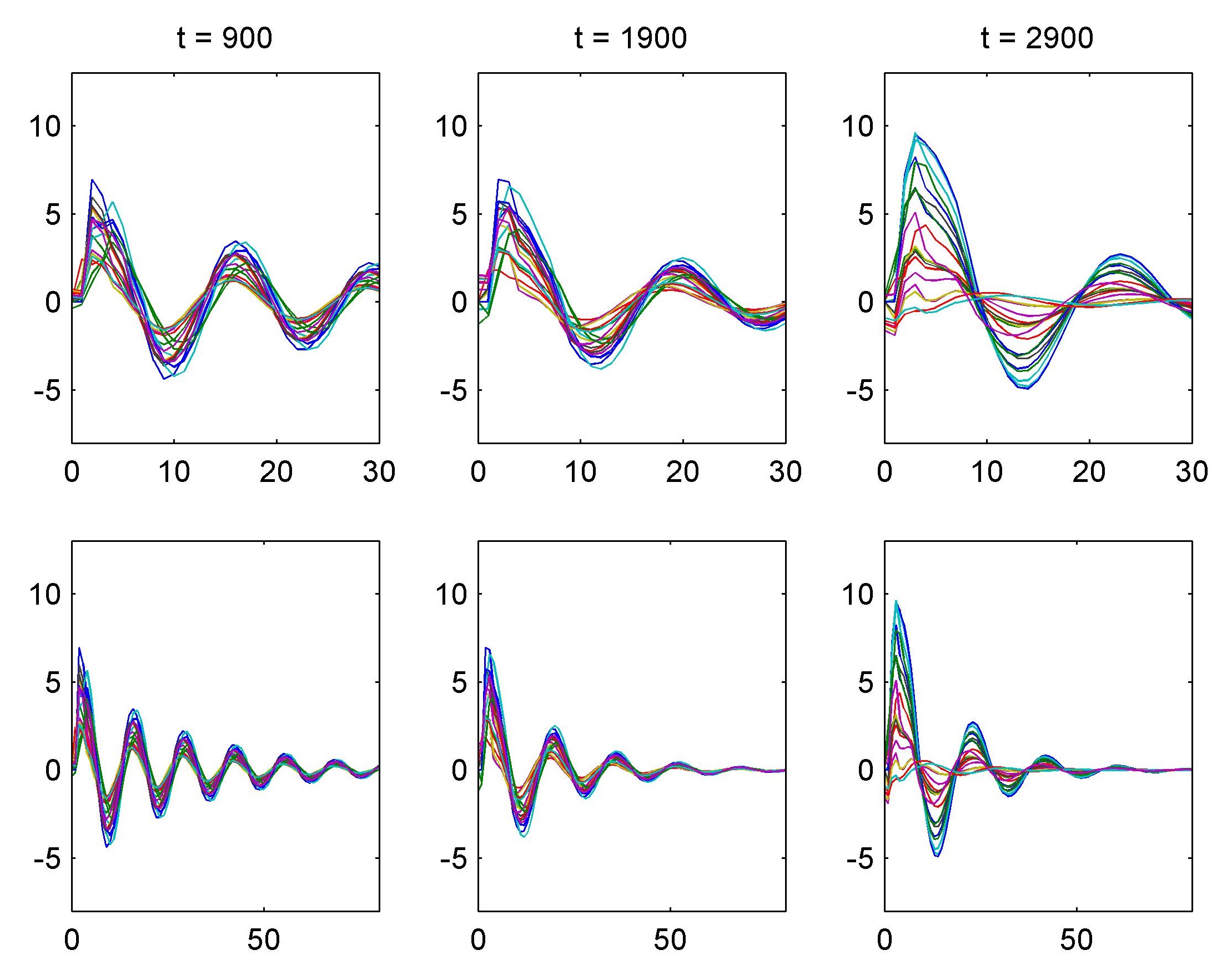} \\[5mm]
(ii) Model M+\\
\includegraphics[width=3.8in]{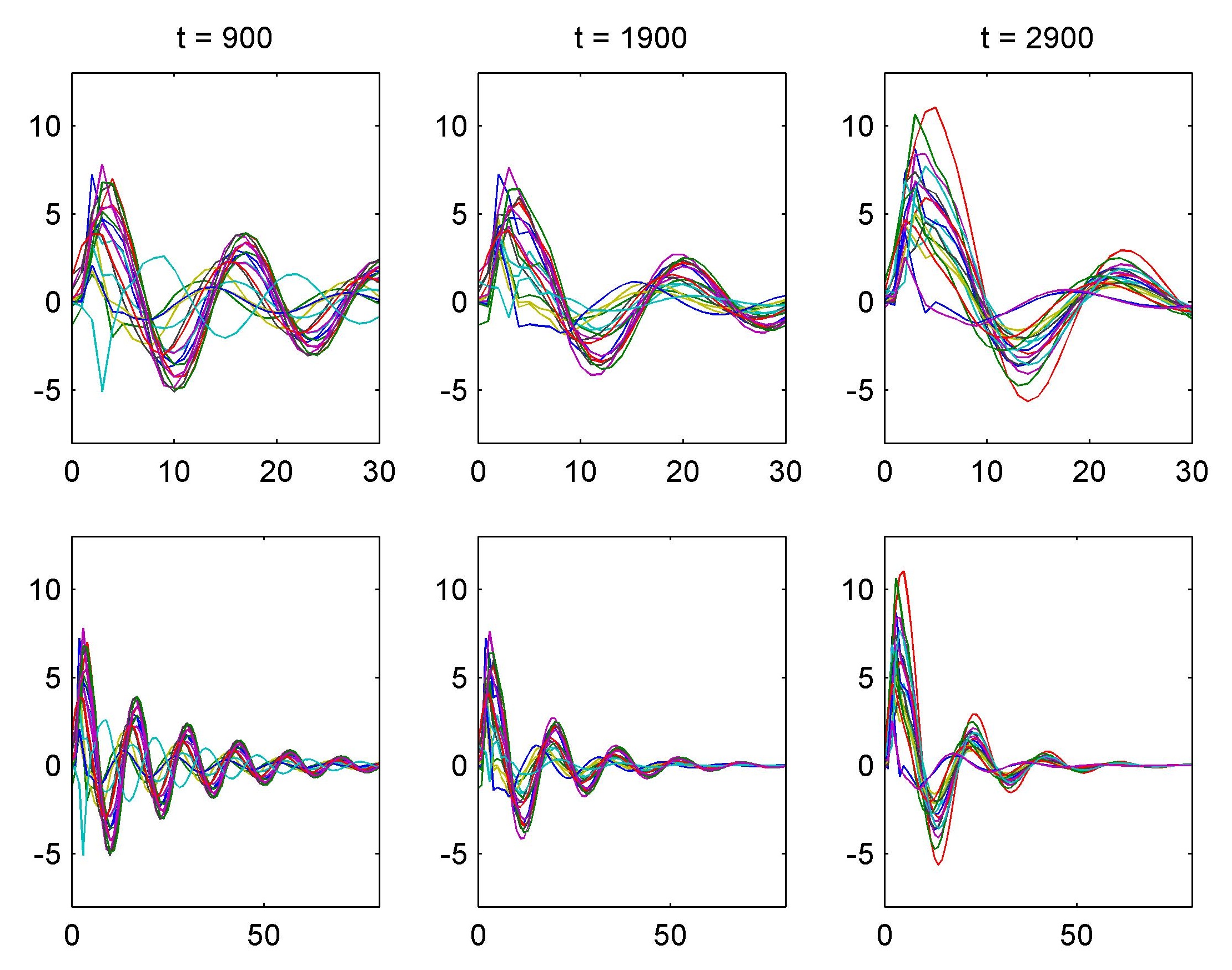}

\caption{EEG analyses: Impulse responses of the $m=19$ EEG channels to a shock to the underlying factor process $x_t$
obtained from (i) Model M,  (ii)  Model M+.   
The impulse response functions computed at three different time points throughout the seizure are shown 
(columns). For each   model,  the impulse response projections are made from the time point indicated by column
header up to 80 time periods ahead (lower rows in (i) and (ii)); the same responses are shown 
on a truncated time period up to only 30 time periods ahead (upper rows in (i) and (ii)), for  clarity.}

\label{fig:imp}

\end{figure}

We do this here from three time points $(t=900, 1900, 2900)$ chosen in the early, middle and later sections of the 
EEG series; this exhibits differences in the model projections/implications over time due to the dynamics, 
as well as differences between the two models in each of these periods.  
Computations are easily done  by using the posterior MCMC samples to project forward in time;
predictive expectations are then computed as  Monte Carlo averages. 
The impulse value $e$ is taken as  the average over $t=1:T$ of the
estimated historical innovations standard deviations   $E(w_t^{1/2}|\y_{1:T})$. 
Figure~\ref{fig:imp} plots the impulse responses of the 19 EEG channels with $h=80$ and from each of the two models. 
Note that, for our comparison purposes here, we are interested in the {\em forms} of the 
impulse responses over the horizon specified, not their specific values.  We already know that the 
innovations variance $w_t$ shows marked changes over time and, in particular, decays to rather low 
values in the later stages of the seizure. Hence the shock size $e$ taken here is larger than relevant and realized
innovations over the latter part of the time period, and  the amplitudes of impulse responses should therefore
not be regarded as pertinent. The form of the projections are the focus.

\begin{figure}[p]
\centering
(i) $h=2$ \\
\includegraphics[width=4in]{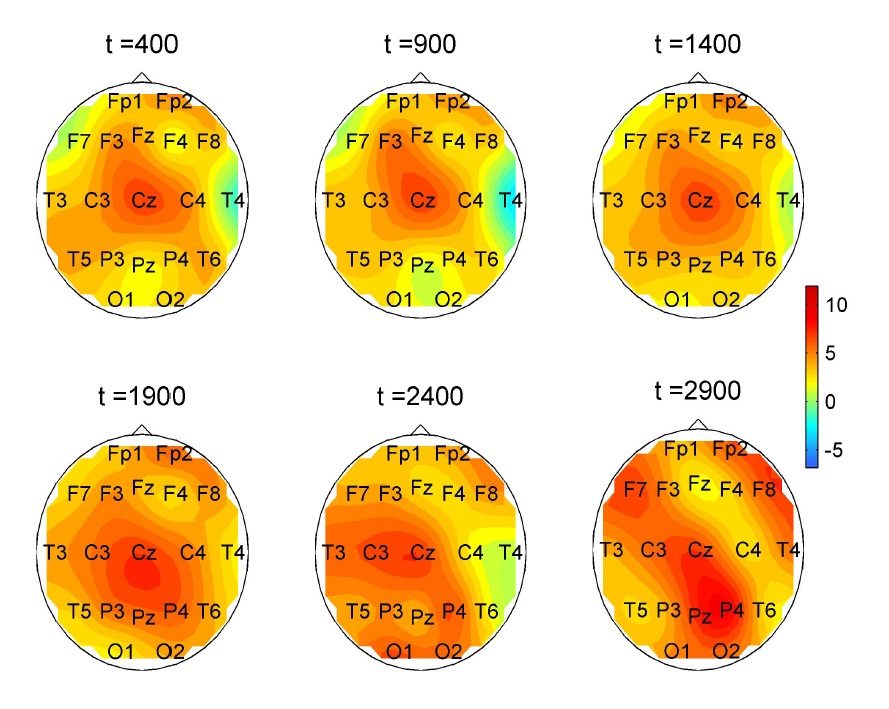} \\[8mm]
(ii) $h=8$ \\
\includegraphics[width=4in]{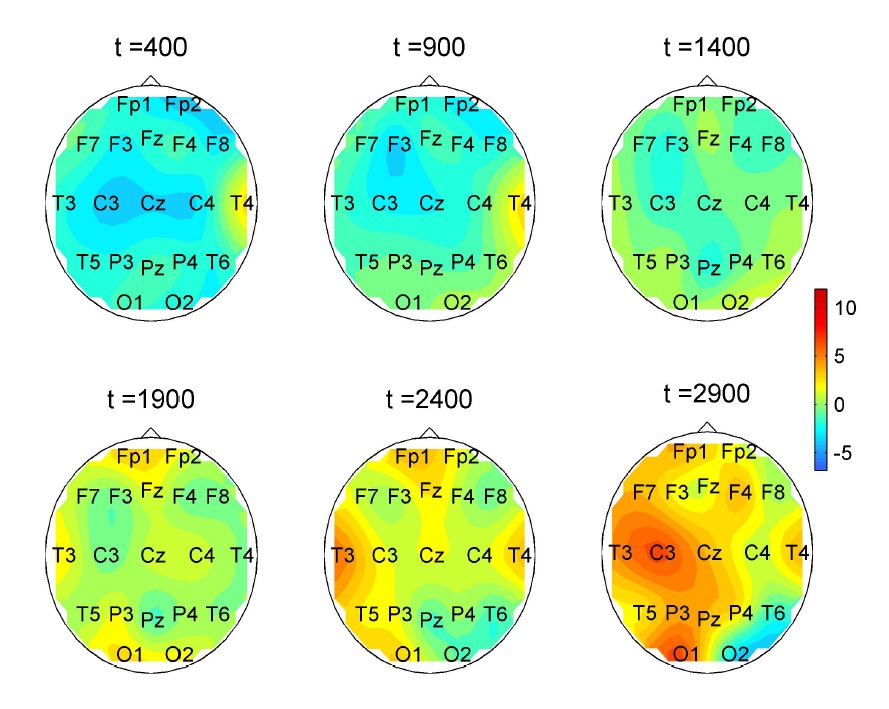}
\caption{EEG analyses: Interpolated patterns of impulse responses from Model M+.   
These are computed at the six time points indicated, and shown for selected impulse response 
horizons $h=2$ and $8$, respectively.}
\label{fig:impulses}
\end{figure}

Patterns of the impulse response are clearly time-varying across the three exhibited time points; variation is 
evident with respect to wave frequency, persistence/decay speed, and variation across the channels. 
In early periods of the seizure, the responses decay slowly with a high-frequency cyclical wave, while in later periods the
decay is more rapid and the oscillations at lower frequency. While there are, as we have discussed above, marked patterns of
variation in lag/lead relationships across channels, there is the appearance of stronger synchronicity earlier in the 
seizure episode, and this deteriorates towards end of the seizure. 

The responses from the TV-VAR extended Model M+ model exhibit more variation across the channels than those from the 
Model M. This is attributable to the induction of some spill-over effects of the shock. Through the latent factor model component alone, 
the shock $\epsilon_{t+1}=e$ 
has an impact on each of the channels through its immediate influence on $x_{t+1}$ and the consequent transfer response
of these effects via $x_{t+1}, $ and so forth. In the extended model, additional feed-forward effects are passed through the channels 
via the TV-VAR component. Some additional insights into the nature of impulse responses can be gained from 
Figure~\ref{fig:impulses} that shows images  interpolating the 9 channel responses across the brain areas, based on 
analysis of the extended Model M+. These are
shown at six time points across the seizure period, and for selected horizons $h=2$ and $8$, respectively; these images 
clearly show the time variation of the responses spreading over the channels. 
 
An animated figure, available as online supplementary material,  
provides a dynamic display over impulse response horizons 1:80, with a movie that 
more vividly  exhibits the differences due to time period.   
The animations 
\href{http://www.stat.duke.edu/~mw/.downloads/NakajimaWest2016EEG/animate-impulse.avi}{(linked at the external site here)}
represent the six time points in Figure~\ref{fig:impulses}, and show images of the
impulse responses as the projections are made over  $t+1,t+2,\ldots,t+h$ to horizon $h=80.$

\section{Concluding Remarks}

The EEG time series analysis highlights the utility of latent thresholding dynamic models in 
constraining model parametrization adaptively in time, with resulting improvements in 
intepretation and inferences on inter-relationships among series and transfer response characteristics. 
An additional comment on model comparison in the case study is worth mentioning. 
Statistical evaluation and comparison of Model M$+$ with Model M is implicit since the latter is a special case of 
the former. The analysis results of M$+$ explicitly show the relevance of the
extensions and hence support the more general model.    This is separately supported by values of the 
deviance information criterion  (DIC; see \citealp{Spiegelhalter02}) computed from the MCMC results for each model
separately; this yields estimated DIC is 996,191.7 for Model M and 988,435.9 for Model M$+$, 
which indicates strong evidence that Model M$+$ dominates Model M.  

A number of methodological and computational areas remain for further study. Among them, we note potential for
integrating spatially-dependent structures with latent threshold factor models, motivated in part by the 
 spatial-temporal findings in the EEG study. Also, incorporating two or more common latent processes might 
 allow evaluation of more complex latent factor structures for these and other applications.  Computational challenges are clear in connection 
 with applying these models to higher dimensional time series such as are becoming increasing common in 
 neuroscience as they are other other areas.  That said, we expect the dynamic  latent thresholding approach
to become increasing relevant and important-- in constraining and reducing effective parameter dimension via dynamic sparsity
 in model parameters-- in contexts with higher-dimensional time series. 

\newpage
 
\section*{Appendix: Summary of MCMC Analysis} 

Based on the observations $\y_{1:T}$, the full set of latent state parameters and model parameters for the posterior analysis of
DTRFM Model M  is as follows: 
\begin{itemize}\itemsep1pt
  \item The latent factor process states $ x_{-p+1:T} $ including uncertain initial values; 
  \item The latent TVAR coefficient process $\bdelta_{1:T} $;
  \item The variance processes $\bSigma_{1:T}$ and $w_{1:T}$;
  \item The latent  factor loading process  $\B_{0:T}$, including the uncertain initial state;
  \item The hyper-parameters $\btheta = \{ \mu_{ik}, \phi_{ik}, v_{ik}; \, i=2:m,\ k=1:r \}$ and $\bPsi$;
  \item The latent threshold hyper-parameters $  d_{2:m,1:r} $.
\end{itemize} 
Key components of the MCMC are below. We simply note the states or parameters being generated, 
implicitly conditional on all other states and parameters.

\smallskip\para{Latent factor process states $ x_{-p+1:T} $}\newline
The model of \eqns{DFM}{delta} can be written in a conditionally linear, 
Gaussian dynamic model form with a modified state $\tilde{\f}_t = (x_t,\ldots,x_{t-p+1})'$ and a state transition
	\begin{eqnarray}
	\label{eq:state_tilde_f}
	\tilde{\f}_t &=& \G_t \tilde{\f}_{t-1} + \e_t	\end{eqnarray}
where 
	\begin{eqnarray}
	\G_t &=& \left(
	\begin{array}{ccccc}
	 \delta_{1t} & \delta_{2t} & \cdots & \delta_{p-1,t} & \delta_{pt} \\
	 1 & 0 & \cdots & 0 & 0 \\
	 0 & 1 & \cdots & 0 & 0 \\
	 \vdots & & \ddots & & \vdots \\
	 0 & 0 & \cdots & 1 & 0
	\end{array}
	\right)\quad\textrm{and}\quad
	\e_t = \left( 
	\begin{array}{c}
	 \varepsilon_t \\ 0 \\ 0 \\ \vdots \\ 0
	\end{array}
	\right).
	\end{eqnarray}
Generation of the full sets of states is obtained by the standard forward filtering, backward sampling (FFBS) algorithm~\citep[e.g.][]{PradoWest10}, which is efficient in the sense that the full trajectories of the states over time are regenerated at each iterate of the overall MCMC.

\smallskip\para{TVAR coefficients}\newline
Conditional on $x_{-p+1:T}$ and the variances $w_{1:T}, \bPsi$, \eqns{xt}{delta} reduce to a univariate, linear and
Gaussian dynamic regression model with respect to the state process $\bdelta_{1:T}$. We sample the states using the FFBS algorithm.

\smallskip\para{TVAR innovations volatility}\newline
Based on the standard inverse gamma/beta Bayesian discount model for the variance sequence $w_{1:T}$ over time as noted in
Section~\ref{subsec:DTRFM},  the corresponding FFBS for volatilities provides a full sample from the 
conditional posterior for  $w_{1:T}$ given all other quantities.  

\smallskip\para{Observation variances}\newline
Similarly, the full conditional  posterior for the $\bSigma_{1:T}$ factorizes into $m$ components involving the 
individual $\sigma_{i,1:T}$ separately over $i,$  and the discount variance FFBS applies to each in parallel
to generate full conditional posterior samples.

\smallskip\para{Factor loading process states}\newline
Following \cite{NakajimaWest10, NakajimaWest11}, we sample each $\bbeta_t=\{\beta_{1:k,t}\}$ from its conditional posterior distribution given $\bbeta_{-t}=\bbeta_{0:T}\backslash\bbeta_t$ and all other parameters. Recall that the elements of 
$\bbeta_t$ follow standard AR(1) processes, but are linked to the observation equation by the latent threshold structure. 
The resulting conditional posterior for $\bbeta_t$ is a non-standard distribution that we cannot directly  sample. We use a Metropolis-within-Gibbs sampling strategy with the proposal distribution derived in the {\it non-threshold} case by assuming $s_{ikt}=1$; i.e., we generate the candidate from a standard linear dynamic model for $\bbeta_t$ without the latent thresholds~\citep[see Section 2.3 of][]{NakajimaWest10}.

\smallskip
\para{Hyper-parameters of AR and TVAR model components}\newline
Priors for the latent AR hyper-parameters $\btheta$ assume prior independence across series $i=2:m$ with traditional forms: 
normal or log-gamma priors for $\mu_{ik}$, truncated normal or shifted beta priors for $\phi_{ik},$ and inverse gamma priors for 
$v_{ik}^2$. On this basis, the full conditional posterior for $\btheta$ breaks down into conditionally independent components across $i=2:m.$ 
We then resample  the $(\mu_{ik}, \phi_{ik}, v_{ik}^2)$ in parallel across $i,$ using direct sampling from the conditional posterior in cases that the
priors are conditionally conjugate,  or alternatively  via Metropolis Hastings steps.  

For the TVAR error variance matrix $\bPsi$, an inverse Wishart prior leads to an easily sampled inverse Wishart complete conditional posterior. 

\smallskip
\para{Latent thresholds hyper-parameters}\newline
As discussed in Section~\ref{subsec:DTRFM}, the structured prior for the thresholds $d_{ik}$ takes them as conditionally independent 
over $i=2:m,\ k=1:r,$ with marginal priors that depend on the parameters of the corresponding latent AR processes, viz. 
 $d_{ik}\sim U(0, |\mu_{ik}| + K u_{ik} )$ where $u_{ik}^2 = v_{ik}^2/(1-\phi_{ik}^2).$    The set of thresholds are then 
also independent in the complete conditional posterior; they are resampled in parallel via Metropolis Hastings independence chain steps 
using the conditional uniform priors as proposals.   This is precisely as pioneered in \cite{NakajimaWest10,NakajimaWest11} in other latent threshold
 models, and its efficacy has been borne out in a number of examples there. 

Finally, note that the above requires a slight modification and extension to generalize the MCMC for the extended DTRFM Model M+
of Section~\ref{subsec:TVVARDTRFM}.
The extension now involves the TV-VAR parameter matrices $\A_{1:T}$ in \eqn{DFM} with $\z_t=\y_{t-1},$ together with the required
latent initial \lq\lq missing'' vector $\y_0.$   The above development applies conditional on these elements with the obvious modifications
to subtract $\A_t\y_{t-1}$ from $\y_t$ throughout.   Then additional MCMC steps are needed. First,  $\y_0$ is generated from a complete conditional
normal posterior under a suitably diffuse normal prior. Second, the latent thresholded elements of the 
sequence $\A_{1:T}, $ and the set of hyper-parameters of the underlying AR(1) processes as well as the corresponding thresholds, 
are treated just as are the elements of $\B_{1:T},$ discussed above. This component is a special case of the MCMC analysis for more general 
TV-VAR models as developed in~\cite{NakajimaWest10}. 

\begin{figure}[h]
\centering
\includegraphics[height=1.8in]{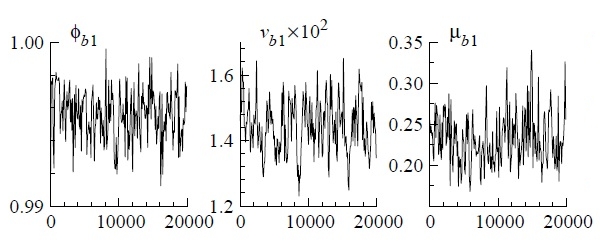} \\
\includegraphics[height=1.8in]{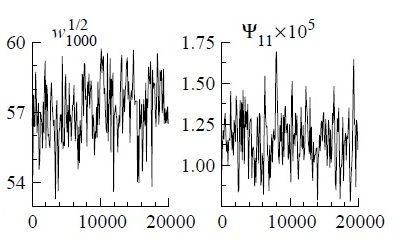} 
\\[8mm]
\caption{
EEG analyses: Typical MCMC trace plots for selected parameters.
}
\label{fig:trace}
\end{figure}

\smallskip
\para{Comment on MCMC convergence}\newline
Some insights into the convergence of the MCMC sampling are gained by viewing trace plots for selected parameters.
As an example,  some such plots from the analysis of the  
extended Model M$+$ are shown in Figure \ref{fig:trace}.


\end{document}